\documentclass[aps,prb,reprint,amsmath,amsfonts,groupedaddress]{revtex4-1}

\usepackage{xcolor}
\definecolor{darkgreen}{rgb}{0,0.6,0}
\definecolor{cyan}{rgb}{0,0.7,0.8}
\usepackage[colorlinks=true,citecolor=darkgreen,urlcolor=cyan]{hyperref}
\usepackage{graphicx}

\newcommand{\mrm}[1]{\mathrm{#1}}

\newcommand{\mbb}[1]{\mathbb{#1}}
\newcommand{\mc}[1]{\mathcal{#1}}

\newcommand{\eref}[1]{(\ref{#1})}
\newcommand{\Eref}[1]{Eq.~(\ref{#1})}

\newcommand{\fref}[1]{Fig.~\ref{#1}}
\newcommand{\Fref}[1]{Figure~\ref{#1}}
\newcommand{\tref}[1]{Table~\ref{#1}}
\newcommand{\sref}[1]{Sec.~\ref{#1}}
\newcommand{\Sref}[1]{Section~\ref{#1}}

\newcommand{\srefs}[2]{Secs.~\ref{#1} and~\ref{#2}}
\newcommand{\aref}[1]{Appendix~\ref{#1}}
\newcommand{\peref}[1]{\protect{(\ref{#1})}}
\newcommand{\pEref}[1]{\protect{Eq.~(\ref{#1})}}
\newcommand{\pfref}[1]{\protect{Fig.~\ref{#1}}}

\newcommand{\ra}{\rangle}
\newcommand{\la}{\langle}

\newcommand{\rcite}[1]{Ref.~\onlinecite{#1}}

\newcommand{\olcite}[1]{\onlinecite{#1}}

\begin{document}

\title{Symmetry-Protected Local Minima in Infinite DMRG}

\author{Robert N. C. Pfeifer}
\email[]{robert.pfeifer@mq.edu.au}
\affiliation{Dept. of Physics \& Astronomy, Macquarie University, Sydney, NSW 2109, Australia}

\date{\today}

\begin{abstract}
The infinite Density Matrix Renormalisation Group (iDMRG) algorithm is a highly successful numerical algorithm for the study of low-dimensional quantum systems, and is also frequently used to initialise the more popular finite DMRG algorithm. Implementations of both finite and infinite DMRG frequently incorporate support for the protection and exploitation of symmetries of the Hamiltonian. In common with other variational tensor network algorithms, convergence of iDMRG to the ground state is not guaranteed, with the risk that the algorithm may become stuck in a local minimum. In this paper I demonstrate the existence of a particularly harmful class of physically irrelevant local minima affecting both iDMRG and to a lesser extent also infinite Time-Evolving Block Decimation (iTEBD), for which the ground state is compatible with the protected symmetries of the Hamiltonian but cannot be reached using the conventional iDMRG or iTEBD algorithms. I describe a modified iDMRG algorithm which evades these local minima, and which also admits a natural interpretation %
on topologically ordered systems with a boundary.
\end{abstract}


\maketitle

\section{Introduction}

The Density Matrix Renormalisation Group (DMRG) algorithm\cite{white1992,white1993,schollwock2005} is a numerical algorithm for computing the ground state of a lattice Hamiltonian on a finite or infinite chain, and is one of the most successful and widespread algorithms in condensed matter physics.\cite{schollwock2011} 
In the last two decades it has also been recognised that both finite and infinite DMRG may be understood as tensor network algorithms, with the ground state being expressed in the form of a Matrix Product State (MPS).\cite{ostlund1995,verstraete2004,mcculloch2007,schollwock2011} 
Implementations of DMRG typically exploit superselection principles with respect to quantum numbers such as particle number or total spin in order to protect these symmetries and/or increase computational efficiency,\cite{mcculloch2000,mcculloch2002,mcculloch2007} and in the tensor network formulation of the DMRG algorithm this equates to working with tensors which explicitly respect and exhibit the corresponding global symmetries of the system.\cite{singh2010,singh2011,pfeifer2011a,singh2012}

It has been known for some time now that in the infinite form of the DMRG algorithm (iDMRG), complications may arise in systems having non-bosonic statistics.
Such systems typically converge to a stable orbit in the space of MPS states rather than to a single state, with the spread of the energies of these states being determined by the truncation accuracy of the MPS.\cite{caprara1997,caprara1997a,mcculloch2000}
In the present paper I demonstrate a more serious problem which may arise in iDMRG simulations with non-trivial particle statistics, if these statistics are also exploited for computational advantage: 
When the MPS tensors of the true ground state act as intertwinors on the space of protected symmetry charge labels with a period $p>2$,
the initial boundary conditions of the simulation may force it to converge to a state which is not (and does not closely approximate) the ground state.

In \sref{sec:iDMRG} of this paper I review the iDMRG algorithm, while in \sref{sec:iDMRGproblem} I describe how the performance of the infinite DMRG algorithm may suffer when symmetries of the Hamiltonian are exploited for computational advantage, and give a simple example in which the MPS tensors of the ground state act as period-3 intertwinors.
In \sref{sec:symIDMRG} I show how the iDMRG algorithm may be modified to circumvent this problem and avoid these local minima, while \sref{sec:anyonIDMRG} discusses how the improved algorithm has a natural physical interpretation when applied to topologically ordered systems with a boundary.\cite{kitaev2006,bonderson2007,bonderson2008,pfeifer2012a} 
A more realistic example is given in \sref{sec:example}, which demonstrates the superior convergence of the revised algorithm. Finally \sref{sec:iTEBD} argues that equivalent local minima are expected to exist for the iTEBD algorithm, %
though this algorithm is seen to %
avoid most symmetry-protected local minima as a result of the boundary conditions implicit when the simulation is initialised.

\section{Infinite DMRG\label{sec:iDMRG}}

\subsection{Overview\label{sec:iDMRGoverview}}

\begin{figure}
\includegraphics[width=\columnwidth]{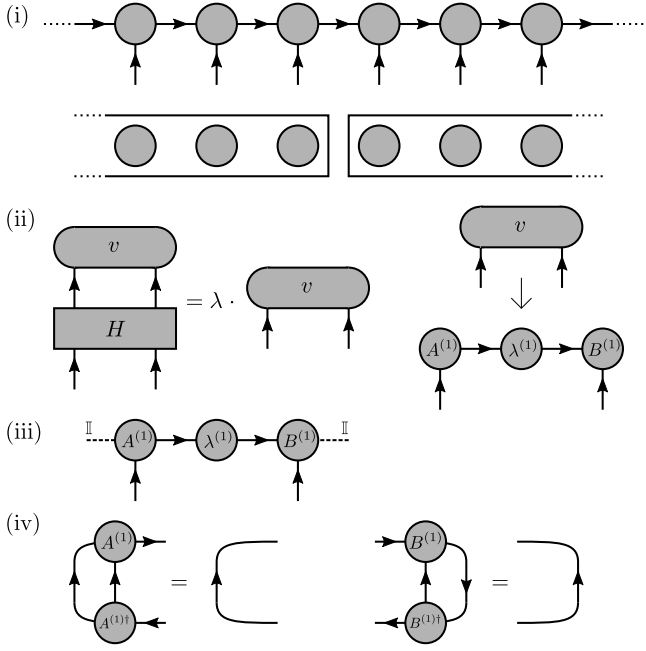}
\caption{(i)~The infinite Matrix Product State, in tensor network and traditional DMRG notations. Orientations are applied to all indices of the MPS to enable the exploitation of quantum numbers which are not self-dual. (ii)~Construction of the initial two-site MPS. Object $v$ is the lowest-energy eigenvector of the two-site Hamiltonian with eigenvalue $E_0$, and is decomposed into three components, $A^{(1)}$, $\lambda^{(1)}$, and $B^{(1)}$, where the superscript $^{(1)}$ indicates that these are the tensors constructed during iteration~1 of the algorithm. %
The $A$ and $B$ tensors typically have three indices apiece, so we may also represent this as shown in diagram~(iii) where the additional indices are of dimension~1 and carry trivial quantum number (also described as ``the vacuum charge'', and frequently denoted $\mbb{I}$). (iv)~All $A$ and $B$ tensors satisfy left-unitarity and right-unitarity respectively. %
\label{fig:iMPS}}
\end{figure}

Behind every tensor network algorithm lies a tensor network Ansatz, and for DMRG, this Ansatz is the MPS. For an infinite system this Ansatz is likewise, in principle, infinite in extent, as per \fref{fig:iMPS}(i). 
In practice the limit of an infinite Ansatz is approached asymptotically. The iDMRG algorithm is initialised by exactly diagonalising a two-site Hamiltonian and decomposing the lowest-energy eigenvector $v$ into three components, $A^{(1)}$, $\lambda^{(1)}$, and $B^{(1)}$, as shown in \fref{fig:iMPS}(ii)-(iii). Note that $A^{(1)}$ and $B^{(1)}$ are typically gauged to be left-unitary and right-unitary respectively, as shown in \fref{fig:iMPS}(iv), and $\lambda^{(1)}$ is diagonal and non-negative. Tensors $A^{(1)}$, $B^{(1)}$, and $\lambda^{(1)}$ thus correspond to the Schmidt (or singular value) decomposition of the eigenvector $v$. Iteration~2 of the algorithm then consists of taking the centre two sites of the Ansatz, inserting an additional two sites in between them, and computing a new triplet of tensors $A^{(2)}$, $\lambda^{(2)}$, and $B^{(2)}$ so as to minimise the total energy of the Hamiltonian (now on four sites), subject to fixed $A^{(1)}$ and $B^{(1)}$ and constraints on the maximum permitted index dimensions of the tensors. Tensors $A^{(2)}$, $B^{(2)}$, and $\lambda^{(2)}$ are gauged as before. 

\begin{figure}
\includegraphics[width=\linewidth]{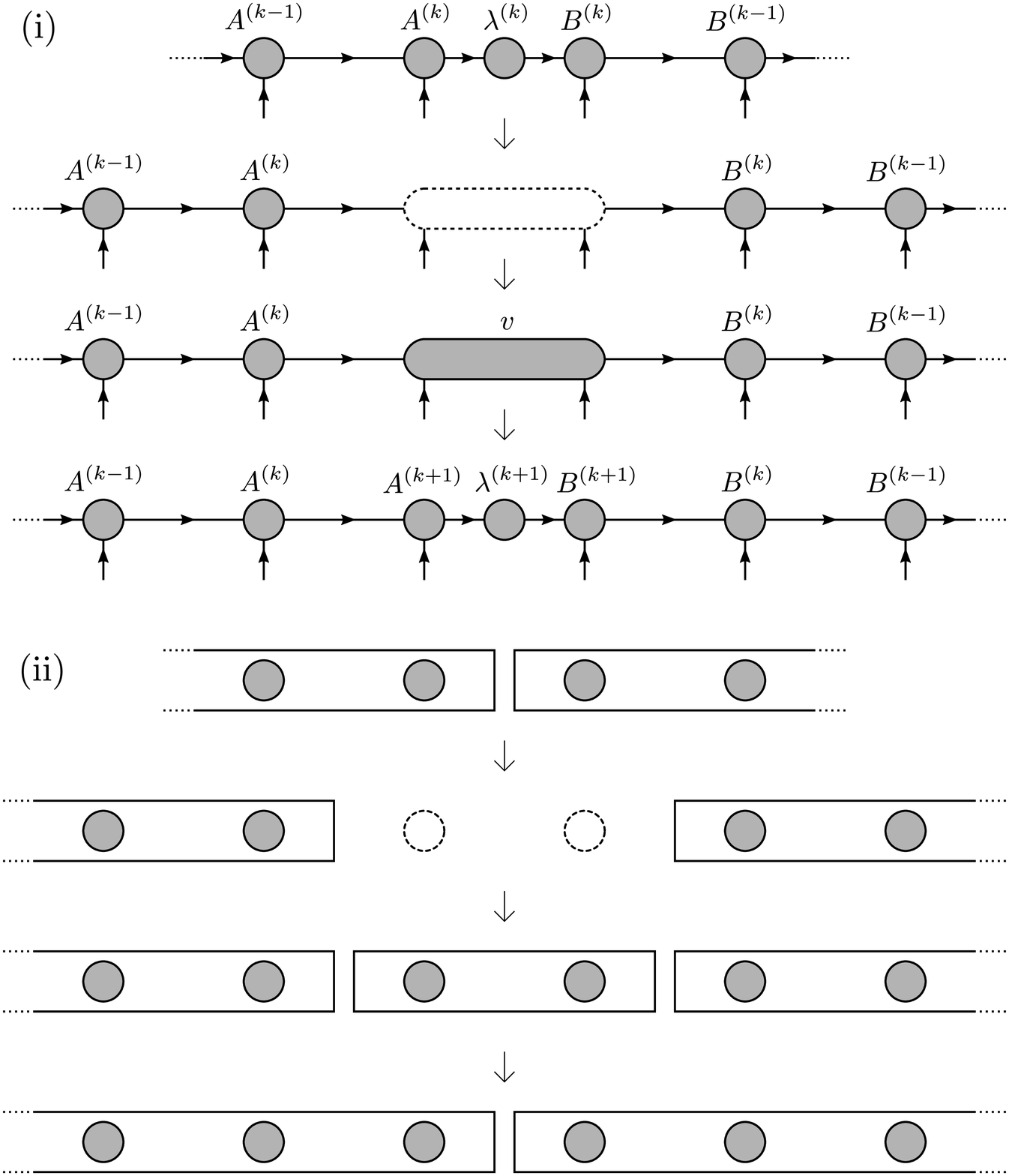}
\caption{One iteration of the infinite DMRG algorithm, (i)~in tensor network notation and (ii)~in traditional DMRG notation.\label{fig:iDMRG}}
\end{figure}

This tensor pair insertion procedure is then iterated repeatedly until a steady state is reached such that $\lambda^{(k)}\approx\lambda^{(k-1)}$ for all $k>k_\mrm{convergence}$, to within some desired tolerance 
\begin{equation}
\sum_i\left({\lambda^{(k)}}^i_i-{\lambda^{(k-1)}}^i_i\right)^2\leq\varepsilon, 
\end{equation}
or that the values of the tensors $\lambda^{(k)}$ are cyclic with some integer period $p$ for all $k>k_\mrm{convergence}$, i.e.~$\lambda^{(k)}\approx\lambda^{(k-p)}$ for all $k$ after convergence has been attained, again within some tolerance $\varepsilon$. Formally, exact convergence corresponds to $\varepsilon\rightarrow 0$, but in practice this is limited by the numeric precision of the software implementation.
This process is illustrated in \fref{fig:iDMRG}.

Note that there exists an additional gauge freedom in the choice of $A^{(i)}$, $\lambda^{(i)}$, and $B^{(i)}$ which is not fixed by the above prescription.
The decomposition 
\begin{equation}
v\rightarrow \{A^{(i)},\lambda^{(i)},B^{(i)}\} 
\end{equation}
is invariant under a transformation of the form
\begin{equation}
\begin{split}
A^{(i)}\rightarrow A^{(i)}U\\
B^{(i)}\rightarrow U^\dagger B^{(i)}
\end{split}\label{eq:Ugauge}
\end{equation}
where $U$ is a block-diagonal unitary matrix acting on the index shared with $\lambda^{(i)}$, whose block sizes correspond to the degeneracy of the associated singular values in $\lambda^{(i)}$. Thus, for example, if all entries in $\lambda^{(i)}$ are unique, then $U$ is a diagonal matrix of complex phases. This gauge may be changed at any time by inserting a pair of matrices $UU^\dagger$ on an internal bond of the MPS and absorbing $U$ and $U^\dagger$ into the $A$ or $B$ tensors to the left and right of the bond.

On convergence, there exists a choice of gauge as per \Eref{eq:Ugauge} such that an infinite MPS may be constructed by infinitely repeating blocks made up of the last $p$ tensors on each side,
\begin{equation}
\lim_{n\rightarrow\infty}\left[A^{(k-p+1)}\ldots A^{(k)}\right]^n \lambda^{(k)}
\left[B^{(k)}\ldots B^{(k-p+1)}\right]^n,\label{eq:blockMPS}
\end{equation}
and optionally absorbing matrix $\lambda^{(k)}$ into either $A^{(k)}$ or $B^{(k)}$ to obtain the form of \fref{fig:iMPS}(i).\footnote{For \protect{$\varepsilon=0$} and infinite MPS bond dimension \protect{$D$} it is in theory possible to construct a Hamiltonian for which \protect{$p$} is infinite; an example would be a system whose filling fraction tends towards an irrational number as \protect{$k\rightarrow\infty$}, with particle number a protected symmetry. By definition such a system does not converge for any finite $k$. In practice, with finite \protect{$D$} and \protect{$\varepsilon$} and a deterministic computing device, the state space of the calculation is finite and any trajectory through this space will eventually repeat for some \protect{$p$}, which may or may not be small enough to be recognised in practice.}
The low-energy state to which the iDMRG algorithm has thus converged is then assumed to closely approximate the ground state of the infinite system.

In order for this assumption to be valid, it is necessary (though not sufficient)
\begin{enumerate}
\item that the ground state of the system may be well-approximated (or exactly represented) by an infinite MPS, and
\item that the iDMRG algorithm is capable of converging to that MPS.
\end{enumerate}
Where multiple degenerate ground states exist which may be written in MPS form, these are associated with a degeneracy of singular values in $\lambda^{(k)}$ and therefore related by a choice of MPS gauge. In such a case it is therefore always possible to write an orthogonal basis for the ground subspace with each basis vector taking the form of \Eref{eq:blockMPS}.

\subsection{A note on notation}

Over the years, a number of different nomenclatures have been adopted for the tensors of the MPS Ansatz. In this paper, when MPS tensors are gauged according to \fref{fig:iDMRG} (left- or right-unitary gauge) they are denoted $A$ or $B$ respectively. Tensors which are not necessarily in the left-unitary or right-unitary gauge have historically been denoted by either $A$ or $\Gamma$; this paper will use $\Gamma$ to avoid ambiguity.

Regarding arrows on the indices of the MPS, these arrows have historically been used for two purposes. They may either signify renormalisation group flow, with the $A$ and $B$ tensors performing a coarse-graining on the Hilbert space of the lattice and $\lambda$ being a matrix representation of a ket acting on a Hilbert space of dimension $D\times D$ (where $D$ is the maximum permitted dimension of the MPS bond indices), or they may indicate an orientation for the summing of charges in simulations with protected symmetries.\cite{singh2010} In the former, it is conventional for all arrows to flow towards $\lambda$ and thus those on the MPS bond indices of the $B$ tensors would point from right to left. The latter imposes no strict constraint on orientation, requiring only that reversal of an arrow be associated with replacing each charge on that index with its dual.

While the $AB\lambda$ form of the MPS Ansatz [e.g.~\fref{fig:iDMRG}(i)] is of great practical use when implementing the iDMRG algorithm, writing an MPS in this form may conceal relevant structure as the Ansatz itself explicitly breaks translation invariance. This drawback is avoided when using the $\Gamma$ form of the Ansatz [e.g.~Figs.~\ref{fig:iMPS}(i) and~\ref{fig:Z3MPS}], in which a state which is invariant under translation by $p$ sites may be represented by an MPS made up by periodically repeating $p$ different tensors $\Gamma_1\ldots\Gamma_p$. As this paper makes use of both the $AB\lambda$ and the $\Gamma$ forms of the Ansatz, a convention is adopted in which all bond indices flow from left to right---both in the $AB\lambda$ form and in the $\Gamma$ form---for greater consistency between the %
two notations.

Finally, a wide variety of terminology is used to describe the different regions of the MPS chain when performing DMRG. Historically, in the $AB\lambda$ form of the Ansatz the $A$ portion of the chain, to the left of $\lambda$ and excluding any new tensors being inserted or updated, has been termed the \emph{system block} while the $B$ portion (to the right of $\lambda$, excluding any new tensors being inserted or updated) has been termed the \emph{environment block}. When the tensors being inserted or updated are included, these regions become the \emph{system superblock} and \emph{environment superblock} respectively. The DMRG algorithm is symmetric with respect to the $A$ and $B$ portions of the chain, and so from an MPS perspective the distinction between the system and the environment is only of historical importance, but the distinction between block and superblock remains useful.

\section{Local minima from symmetry protection\label{sec:iDMRGproblem}}

\subsection{Illustrative example\label{sec:illus}}

As a simple example of where the second assumption of \sref{sec:iDMRGoverview} fails, and the iDMRG algorithm is incapable converging to the true ground state of the system, consider the Hamiltonian
\begin{equation}
\hat H=-\sum_i \mu \hat n_i\label{eq:muH}
\end{equation}
on a chain populated by hard-core particles, and introduce a conserved quantum number corresponding to particle number modulo~3. This may seem artificial in the context of \Eref{eq:muH}, but arises naturally when a chemical potential having the form of \Eref{eq:muH} is attached to a pre-existing Hamiltonian having $\mbb{Z}_3$ symmetry, in the limit that $\mu$ becomes large (see e.g.~\sref{sec:example} and \rcite{pfeifer2015a}).

\begin{figure}
\includegraphics[width=\columnwidth]{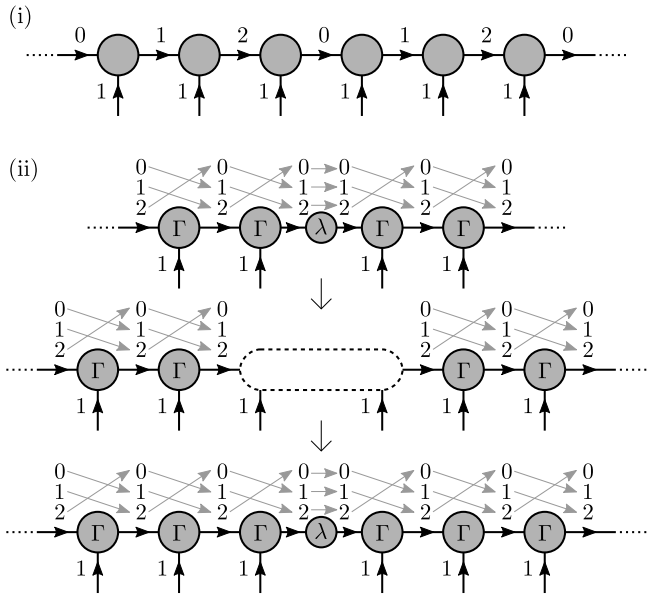}
\caption{(i)~Ground state of \pEref{eq:muH} with three-site translation invariance, requiring MPS bond index dimension $D=1$. %
Index labels correspond to charge sectors of non-zero dimension. (ii)~Insertion of two extra sites into $|\psi'\ra$ of \pEref{eq:transinv}. Grey arrows show how the tensor $\Gamma$ acts as an example on the charge sectors of the MPS bond.\label{fig:Z3MPS}}
\end{figure}

For $\mu>0$, the Hamiltonian \eref{eq:muH} has a unique ground state corresponding to complete filling of the chain. This state admits an MPS representation consistent with protection of the $\mbb{Z}_3$ symmetry, as shown in \fref{fig:Z3MPS}(i), and has bond dimension $D=1$.

In this representation, if the MPS bond index between sites $i$ and $i+1$ carries charge~0, then the index between $i+1$ and $i+2$ carries charge~1, and more generally, that between $i+q$ and $i+q+1$ carries charge~$(q~\mrm{mod}~3)$. This MPS is not translation-invariant, but we may make it so by defining $\hat T$ to be the one-site translation operator and writing
\begin{equation}
|\psi'\ra=\frac{1}{\sqrt{3}}\left(|\psi\ra+\hat T|\psi\ra+\hat T^2|\psi\ra\right),\label{eq:transinv}
\end{equation}
with MPS bond dimension $D=3$. The factors of $1/\sqrt{3}$ are incorporated by arbitrarily selecting a single MPS bond and introducing an additional tensor on this bond, $\lambda=\mrm{diag}(1/\sqrt{3},1/\sqrt{3},1/\sqrt{3})$. While it would be conventional to denote tensors to the left of this bond by $A^{(k)}$, $A^{(k-1)}$, \ldots, and tensors to the right by $B^{(k)}$, $B^{(k-1)}$, \ldots, all of these tensors are in fact identical. In \fref{fig:Z3MPS} they are therefore all denoted $\Gamma$ to emphasise this fact.
The new state is invariant under the operation
\begin{equation}
\lambda\rightarrow\Gamma\lambda\Gamma,
\end{equation}
which inserts an additional two sites (but leaves the MPS unchanged, as it is infinite). This is represented graphically in \fref{fig:Z3MPS}(ii). State $|\psi'\ra$ therefore in principle represents a possible steady state of the infinite DMRG algorithm. In practice, however, this state cannot be attained by the algorithm in its usual form.

This is readily seen by considering \fref{fig:iMPS}(iii). On this first iteration, the vacuum charge boundary conditions may only be satisfied if the number of particles on the sites is a multiple of three. The hard-core constraint sets an upper bound of two, and thus the Ansatz is forced into a zero-particle state. The bonds between $A^{(1)}$, $\lambda^{(1)}$, and $B^{(1)}$ in \fref{fig:iMPS}(iii) are therefore also in the vacuum charge sector, and this constraint propagates to the next iteration. At each step it therefore remains impossible to break out of the zero-particle sector. On the space of states which can be explored by the iDMRG algorithm with the conventional choice of boundary conditions, this state consequently represents a ``local'' minimum which is not the ground state of the Hamiltonian \eref{eq:muH}.\footnote{The notion of locality on the space of matrix product states is not, in general, rigorously defined. A state \protect{$|\psi'\ra$} may be thought of as local with respect to a state \protect{$|\psi\ra$}, a Hamiltonian \protect{$\hat{H}$}, and an algorithm \protect{$\mc{A}$} if a single iteration of algorithm \protect{$\mc{A}$} with Hamiltonian \protect{$\hat{H}$} is capable of taking \protect{$|\psi\ra$} into \protect{$|\psi'\ra$}.}

Note that if the $\mbb{Z}_3$ symmetry is not enforced, then the iDMRG algorithm does not become trapped in this minimum. Enforcing symmetries reduces the space of possible states to which the algorithm may evolve in one iteration. If (as here) this restriction directly results in the algorithm becoming stuck in a state which is not the true ground state, this may be termed a \emph{symmetry-protected local minimum}. In this instance the local minimum is also associated with complete filling of the lattice, but this is not a prerequisite and in \sref{sec:example} we shall see a more realistic Hamiltonian which also exhibits symmetry-protected local minima at large but incommensurate fillings.

\subsection{MPS tensors as intertwinors\label{sec:intertwinors}}

To see how the tensors $\Gamma$ in \fref{fig:Z3MPS} may be understood as intertwinors, note that under Hamiltonian~\eref{eq:muH} with $\mu>0$ the probability amplitude of any site of the lattice being unoccupied is zero. We may therefore project the physical indices of these tensors into the dimension-1 subspace corresponding to the presence of a particle%
, and the MPS tensors then reduce to matrices. Conservation of charge requires that these matrices act as an intertwinor %
on the charge sectors of the MPS bond index.

\begin{figure}
\includegraphics[width=\linewidth]{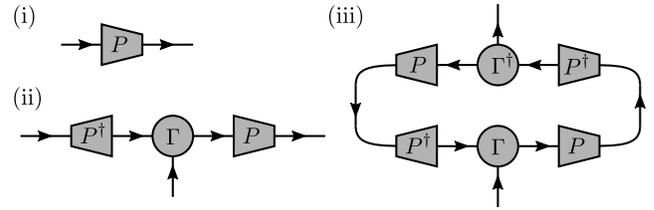}
\caption{(i)~Tensor~$P$ is a symmetry-conserving projection operator acting on the MPS bond index, reducing the dimension of the associated Hilbert space from $D$ to $D'$. (ii)~Tensor~$P$ acts on an MPS tensor $\Gamma$. (iii)~Calculation of the unnormalised 1-site reduced density matrix associated with $P$, denoted $\hat{\rho}^{(P)}$.\label{fig:intertwinor}}
\end{figure}

More generally, let $\Gamma$ be a translation-invariant representation of the ground state of a translation-invariant Hamiltonian $\hat H$,\footnote{If \protect{$\hat H$} is invariant under translation by \protect{$p$} sites, a translation-invariant representation of \protect{$\hat H$} may be constructed by fusing \protect{$p$} sites of dimension \protect{$d$} into one effective site of dimension \protect{$d^p$}.} let $P$ be a symmetry-conserving projector acting on the MPS bond index which reduces the dimension of this index from $D$ to some smaller value $D'$, and let $\mc{Q} = \{q_P\}$ be the space of charges appearing on the smaller index of $P$. Let
$\hat{\rho}^{(P)}$ be the unnormalised 1-site reduced density matrix constructed from the projected MPS tensor $P^\dagger\Gamma_iP$, where $i$ is the physical index, as shown in \fref{fig:intertwinor}. 
Suppose there exists a choice of $P$ such that $\left[\hat{\rho}^{(P)}\right]^i_i$ (no sum over $i$) is non-vanishing for only one value of $i$, denoted $i_P$, and that value $i_P$ is associated with a charge having non-trivial action on $\mc{Q}$.
It then follows that $P^\dagger\Gamma_{i_P}P$ is an intertwinor on the space of charges $\mc{Q}$.
If there exists a set of such projectors $\mc{P}$ such that $\mrm{Tr}(PP^{\prime\dagger})=0$ for all pairs $(P,P')\in\mc{P}$, and
for which the value of 
\begin{equation}
\hat{\rho}^{(\mc{P})} = \sum_{P\in\mc{P}}\left[\hat{\rho}^{(P)}\right]^{i_P}_{i_P}\label{eq:rhoP}
\end{equation}
is sufficiently greater than zero, then this indicates that 
\begin{equation}
\bigoplus_{P\in\mc{P}}\left[\hat{\rho}^{(P)}\right]^{i_P}_{i_P}
\end{equation}
makes a significant contribution to the one-site reduced density matrix, and the object
\begin{equation}
\bigoplus_{P\in\mc{P}} P^\dagger\Gamma_{i_P}P,\label{eq:sumintertwinor}
\end{equation}
which is an intertwinor on a subspace of the MPS bond index, makes a significant contribution to the construction of the ground state. We may then anticipate the existence of a symmetry-protected local minimum. %

Having identified the conditions which may cause the iDMRG algorithm to become stuck in a symmetry-protected local minimum,
the following Sections explore different strategies for modifying (i)~the initial conditions (\sref{sec:extendedv}), and (ii)~the iDMRG update algorithm (\srefs{sec:blocksize}{sec:symIDMRG}), in order to either avoid falling into, or permit escape from, the vacuum sector. In the former approach, the zero-particle state with trivial charges on the MPS bond indices remains a local minimum as once the MPS bond enters this charge configuration, the update algorithm is unable to escape. The latter approach is more robust, eliminating the local minimum from the landscape of states by providing a means for the algorithm to evolve to a lower-energy state, while nevertheless retaining protection of the global $\mbb{Z}_3$ symmetry.

\subsection{Infinite DMRG with free boundary charges\label{sec:extendedv}}

\begin{figure}
\includegraphics[width=\columnwidth]{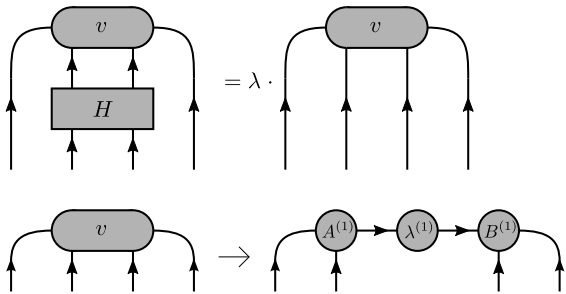}
\caption{Calculation of the initial value of~$v$ with multiple boundary charges. The left index is permitted to take $n_q^{(L)}$ different charges, each of degeneracy~1, and the right index is permitted to take $n_q^{(R)}$ different charges, again each with degeneracy~1. These need not be the same charges as appear on the left. This diagram reduces to \pfref{fig:iMPS}(ii)-(iii) for $n_q^{(L)}=n_q^{(R)}=1$ and $q_L=q_R=\mbb{I}$.\label{fig:extendedv}} %
\end{figure}

One way in which we can avoid the local minima is by allowing multiple choices of boundary charge on the initial state. Object $v$ continues to satisfy $vH=\lambda H$ for the initial 2-site Hamiltonian, but this expression now takes the form shown in \fref{fig:extendedv}.
For the example of \sref{sec:illus}, allowing both the left and the right boundary charges to take any value from $\{0,1,2\}$ results in the iDMRG algorithm finding the translation-invariant ground state $|\psi'\ra$.

It is interesting to compare this approach with the technique used in finite DMRG to compute the lowest-energy state in a given charge sector.\cite{mcculloch1999,mcculloch2007} If we denote the MPS tensors of the finite chain by $\Gamma^{(1)}\ldots\Gamma^{(L)}$,\cite{vidal2004} then the MPS tensors at the end of the chain [either $\Gamma^{(1)}$ or $\Gamma^{(L)}$] customarily have only two indices, and the state computed by the algorithm is the lowest state having trivial quantum number (i.e.~the total charge on all sites is the vacuum). To compute the lowest-energy state in a sector with non-trivial quantum number, one introduces an additional index on either $\Gamma^{(1)}$ or $\Gamma^{(L)}$, and thus on either the system or the environment block. This index behaves similarly to the additional indices introduced in \fref{fig:extendedv}, with the difference that only one such index is introduced, and this index has degeneracy~1.

While the method illustrated in \fref{fig:extendedv} is capable of finding the translation-invariant state $|\psi'\ra$ of \Eref{eq:transinv}, it is still less than ideal. State $|\psi'\ra$ is a ground state with block length~1 in \Eref{eq:blockMPS}, but requires $D\geq 3$. In contrast, state $|\psi\ra$ in \fref{fig:Z3MPS}(i) is a ground state with block length~3, but can be constructed for the lower cost of $D=1$. More generally, if a ground state may be constructed with period $p$ and bond dimension $D$, then there also exists a ground state with period~1 and bond dimension $pD$ with construction analogous to $|\psi'\ra$ in \Eref{eq:transinv}. The approach illustrated in \fref{fig:extendedv} favours construction of solutions in which the charge configuration of the MPS bond has period~1, and thus to obtain the same level of accuracy as a solution with period $p$ in the charge configuration requires a bond index larger by a factor of $p$, increasing computational cost by a factor of $p^3$. This factor may become quite large.\footnote{Note that the algorithm as a whole may still display periodicity---it is only periodicity of charge configurations which is suppressed and which causes increased computational cost. For example, in Refs.~\protect{\onlinecite{caprara1997}}--\protect{\onlinecite{caprara1997a}} conservation of particle number is not exploited for computational gain. The results are asymptotically \protect{$p$}-periodic with \protect{$p>1$}, but all indices carry only the trivial charge.}

The potential weakness of this algorithm is the requirement that the initial boundary charges associated with non-zero entries in $\lambda$, which are selected on a finite system of length~2, should also be appropriate on systems of all lengths $L\in\{2n~|~n\in\mbb{Z}^+\}$. The boundary charges for each iteration are wholly determined by the charges appearing on the MPS bond index of the iteration before, and if early iterations exclude key charges from the bond indices (either because of overly-restrictive bond dimension or because of qualitatively different behaviour in the small-$L$ and large-$L$ regimes) then this may cause failure to converge to the correct ground state. In a worst-case scenario the bond indices may collapse to a single charge sector on some iteration $k$, and the free boundary charge approach then offers no advantage over standard iDMRG with an appropriate choice of total charge. %

An example of failure due to insufficient bond dimension $D$ may be seen by setting $D<3$ with Hamiltonian~\eref{eq:muH}, prohibiting the construction of the translation-invariant state~\eref{eq:transinv}. A more detailed example is provided in \sref{sec:example}.

\subsection{Increasing the inserted block size\label{sec:blocksize}}

Another alternative, suggested in \rcite{mcculloch2000} in the context of protected symmetries associated with particle number, is to modify the size of the initialisation block and subsequent inserted blocks such that the total length of the growing lattice is always an exact multiple of the filling fraction. This approach requires the filling fraction (or whatever other symmetric property is responsible for the charge periodicity) to be known beforehand, and ideally to be a constrained parameter. It therefore cannot be applied to Hamiltonians such as %
\begin{equation}
\hat H=-J\sum_i \left(c^\dagger_i c_{i+1}-c^\dagger_{i+1} c_i\right)-\mu \sum_i c^\dagger_i c_i,
\end{equation}
which is 
the free fermion Hamiltonian written with a chemical potential $\mu$, 
as the filling fraction for such a Hamiltonian is determined during the optimisation process as a result of competition between the chemical potential and the hopping terms.
When compared to the standard approach in which a pair of sites is introduced on each iteration, the additional computational cost of this approach is a factor of ${O}(d^{p-2})$ where $d$ is the dimension of the Hilbert space of a single site.

\subsection{Infinite DMRG with an accessory charge\label{sec:symIDMRG}}

\subsubsection{Description of algorithm}

\begin{figure}
\includegraphics[width=\columnwidth]{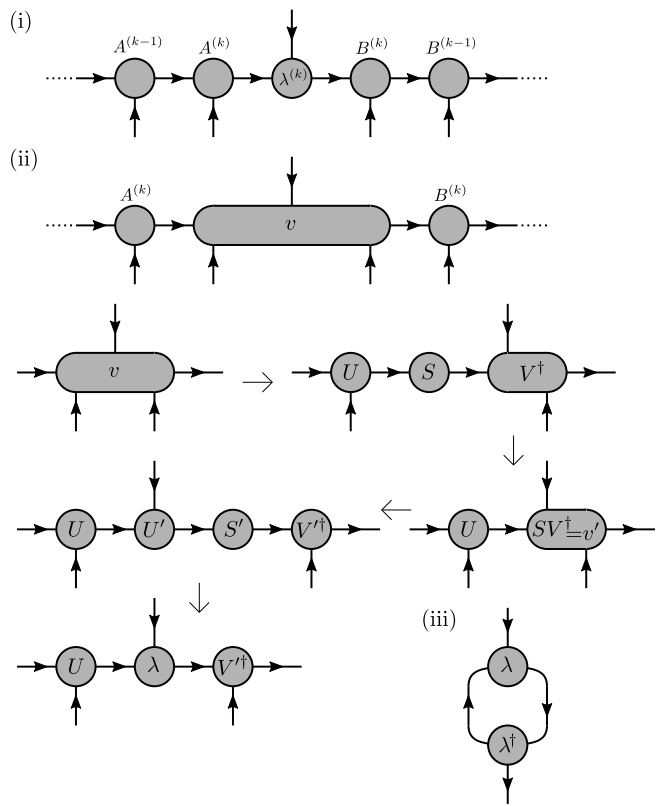}
\caption{(i)~Infinite MPS with an accessory charge. (ii)~Truncation of singular values with an accessory charge. After obtaining the new eigentensor $v$, an initial decomposition divides $v$ into $U$, $S$, and $V^\dagger$. The first truncation is performed over the singular values in $S$. Tensor $v'$ is then constructed from $S$ and $V^\dagger$, then this is decomposed into $U'$, $S'$, and $V'^\dagger$ and a second truncation is performed, over the singular values in $S'$. Tensor $\lambda^{(k+1)}$ is constructed from $U'S'$. (iii)~Calculation of charge sector weights.\label{fig:accMPS}} %
\end{figure}

In \sref{sec:extendedv} the local minimum of Hamiltonian~\eref{eq:muH} corresponding to the empty lattice was avoided by modifying the boundary conditions of the initialisation procedure. \Sref{sec:blocksize} then considered a modification of the iDMRG algorithm which can achieve the same effect, but is possible only under limited conditions. An improvement over both of the above methods may be achieved by modifying the MPS Ansatz as shown in \fref{fig:accMPS}(i).

In this Figure an additional index has been added to the central tensor, $\lambda^{(k)}$, which does not couple to the Hamiltonian and which carries some number of charges $n_q^{(\lambda)}$, each with degeneracy~1. Optimisation proceeds as before, inserting additional sites and computing a new eigentensor $v$ which decomposes into new $A^{(k+1)}$, $B^{(k+1)}$, and $\lambda^{(k+1)}$, only now the $\lambda$-tensor has three indices instead of two. Truncation is performed independently on the left and right sides of $\lambda^{(k+1)}$ during the two singular value decompositions shown in \fref{fig:accMPS}(ii). The cost of the algorithm is increased relative to the customary iDMRG algorithm, but only by a factor of $\mrm{O}[n_q^{(\lambda)}]$.

The additional index on the $\lambda$-tensor may be given a physical interpretation as an accessory site, decoupled from the Hamiltonian, which is capable of carrying arbitrary charge. As with the extra charges introduced at the left and right of \fref{fig:extendedv}, the presence of this charge relaxes the requirement that the total charges on the sites of the left half-lattice and right half-lattice sum to the vacuum sector. %

Since the accessory site is now included on the tensor $v$, which is variationally optimised during each iteration, this approach avoids the problem of locking in choices of boundary charges during initialisation, which may or may not be appropriate in the large-$L$ limit. This algorithm therefore offers greater flexibility than the free boundary charges algorithm of \sref{sec:extendedv}.

\subsubsection{Comparison with targetting of individual charge sectors}

It is useful to compare the accessory site algorithm with the approach discussed in \sref{sec:extendedv} for targetting individual charge sectors in finite DMRG. In both techniques a single additional index is added to the MPS Ansatz, lifting the constraint that the state as a whole must have a quantum number of zero.

When computing the lowest-energy state in a specific charge sector, the additional charge index is customarily placed on either $\Gamma^{(1)}$ or $\Gamma^{(L)}$ and thus is attached to either the system or the environment block. Iteration of the iDMRG algorithm proceeds in the normal fashion, with the total charge constraint imposed by this additional index yielding a state in the desired charge sector. One may target the lowest-energy state across multiple charge sectors simultaneously by admitting several different charges on this additional index. If the additional index is traced over, then the symmetrised version of this process (where extra indices appear on both the system and the environment) is exactly the procedure described in \sref{sec:extendedv}. If the additional index is not traced over, then this is instead equivalent to performing several simultaneous MPS simulations in parallel, looking for the lowest-energy state in each charge sector (henceforth the \emph{multiple charges method}), with the total bond dimension $D$ being divided either statically or dynamically between these simulations. The fraction of $D$ attributed to a given charge sector is then equivalent to the bond dimension of the effective independent MPS on that charge sector. %

The accessory site form of the iDMRG algorithm differs from this approach by putting the extra index on the central tensor $\lambda$, rather than on the system or environment. This simple modification has a number of very important effects.
\begin{enumerate}
\item In the multiple charges method, superselection requires that for an MPS bond dimension $D$, the DMRG algorithm must divide this bond dimension between the multiple different charge sectors being independently optimised. In contrast, accessory site iDMRG typically selects a single charge sector yielding the lowest-energy state and allocates the entire MPS bond dimension to that state.\footnote{A superposition of charge sectors may be selected if both yield equally low energy states, and \protect{$D$} is redundantly large.} For conventional iDMRG this would result in the MPS becoming locked into a single charge sector. This does not happen for accessory site iDMRG, however, because of the second major difference between the multiple charges method and the accessory site method:
\item In the multiple charges method, each of the parallel MPS simulations evolves independently. Thus for a simulation with total charge $q$, the system and environment blocks at step $k$ correspond to the optimal system and environment superblocks \emph{for total charge $q$} at step $k-1$. In accessory site iDMRG, the system and environment blocks at step $k$ correspond to the optimal system and environment superblocks \emph{across all charges} at step $k-1$, even if the optimal state at step $k$ has total charge $q$ and the optimal state at step $k-1$ has total charge $q'\not=q$.
\end{enumerate}
As a result, accessory site iDMRG is capable both of converging more rapidly to a lower-energy configuration than the multiple charges method (as is clearly seen in the example in \sref{sec:example}), and of yielding a more faithful approximation to the true ground state (corresponding to a higher effective bond dimension $D$ in the optimal charge sector at each iteration).

\subsubsection{Interpretation of results}

To interpret the results obtained using this Ansatz, it is necessary to compute the relative weights of the different charge sectors on the accessory site for several iterations. These are given by the diagonal elements of the matrix 
\begin{equation}
M^q_{q'}=\lambda^{\alpha q}_\beta\lambda^{\dagger \beta}_{\phantom{\dagger}\alpha q'} 
\end{equation}
[\fref{fig:accMPS}(iii)]. The behaviour of these values then falls into one of the following four categories:
\begin{enumerate}
\item \emph{Single charge sector}---Only one charge sector has a weight differing from zero by more than can be attributed to machine precision rounding errors, and this sector is consistent from iteration to iteration. The lowest-energy state identified after $k$ iterations takes the form %
of \Eref{eq:blockMPS}, with possible periodicity as per Refs.~\olcite{caprara1997}--\olcite{caprara1997a} and $\lambda^{(k)}$ as per \fref{fig:accMPS}(i). The tensor $\lambda^{(k)}$ carries the specified charge $q$ on the accessory site. If this charge is not the vacuum charge, consider the meaning of this accessory charge in the context of your model. If you wish to find the next-lowest-energy charge sector, repeat the simulation but exclude this charge from the new index of the $\lambda$-tensor. Note that if the only non-trivial sector has the vacuum charge on the accessory site, this site can be projected down to the vacuum sector and then deleted to recover a standard MPS without an accessory site.
\item \emph{Superposition of charge sectors}---When more than one charge sector has a weight different to zero, this indicates the existence of multiple degenerate ground states with different charges on the accessory site. As the ground states are degenerate, the relative occupancies of the charge sectors may vary from iteration to iteration even while the MPS is otherwise converged, unless fixed by a choice of gauge. If wishing to restrict to a single charge value on the accessory site, either project onto this sector and rescale the relevant singular values, or exclude all but one of these sectors and repeat the simulation. The absence of a superposition does not confirm the absence of a degeneracy.
\label{outcome2}
\item \emph{Cyclic charge sectors}---Only one charge sector has a weight different to zero on any given iteration, but this charge changes cyclically from iteration to iteration with a period~$p_q$. To identify a ground state with charge $q$ on the accessory site, halt the simulation when only sector~$q$ has non-trivial weight. The Ansatz for the ground state is now given by %
\Eref{eq:blockMPS} with a block length $p$ which is an integer multiple of $p_q$, being the lowest common multiple of $p_q$ and any periodicity of the sort described in Refs.~\olcite{caprara1997}--\olcite{caprara1997a}. Again note that if the accessory charge is trivial, the accessory site can be projected then deleted.\label{outcome3}
\item \emph{Cyclic superpositions of charge sectors}---A combination of outcomes~\ref{outcome2} and~\ref{outcome3}. 
\end{enumerate}
With limited exceptions (see e.g.~\sref{sec:anyonIDMRG}), the desired outcome will be an Ansatz for a state with vacuum charge on the accessory site. However, by permitting nontrivial charges on the accessory site during the optimisation process one may obtain states such as that of \fref{fig:Z3MPS}(i) which cannot be reached by the usual symmetric iDMRG algorithm. For example, given Hamiltonian~\eref{eq:muH} and $\mbb{Z}_3$ symmetry one obtains a result with cyclic charge sectors (outcome~\ref{outcome3}). Halting when the non-zero weight is in the vacuum sector, deleting the accessory site, and applying \Eref{eq:blockMPS} with block length $p=p_q$ gives state $|\psi\ra$ as shown in \fref{fig:Z3MPS}(i), with MPS bond dimension $D=1$.

Note---A procedure for generating an initial estimate for vector $v$ in conventional iDMRG is given in \rcite{mcculloch2008}. This procedure converges to the same fixed point as the Product Wave-Function Renormalisation Group technique (PWFRG) for TEBD,\cite{nishino1995,hieida1997,okunishi1999,ueda2006} and may be readily generalised to the accessory site algorithm (see \aref{sec:initialvec}). However, it is unable to allow for situations where the charge on the accessory site is cyclic. When this is the case, the overlap of the initial guess with the true lowest-energy eigenvector may be sufficiently small as to result in failure of the numerical algorithm (e.g.~Lanczos) used to find $v$. For this reason, the use of this estimation procedure is not recommended with the current algorithm unless the charge on the accessory site is fixed by the specification of the system to be studied.

\subsection{Topologically ordered systems\label{sec:anyonIDMRG}}

Occasionally, situations may be encountered where it is appropriate to apply an explicit physical interpretation to the accessory site. An example of this is in the study of topologically ordered systems with a boundary. Models of this sort include Kitaev's toric code Hamiltonian\cite{kitaev2003} on a cylinder, critical models whose infra-red limit corresponds to a topological quantum field theory (TQFT),\cite{difrancesco1997,pfeifer2009} %
and models whose particle excitations explicitly correspond to the charges of a TQFT or unitary braided tensor category.\cite{pfeifer2010,konig2010,pfeifer2012,pfeifer2012a}

These models admit dual descriptions either in terms of the microscopic degrees of freedom, such as the spins of Kitaev's toric code, or in terms of the emergent topological order. For the toric code, stars and plaquettes which are not in a lowest-energy configuration may be associated with $e$ and $m$ charges respectively, and may be created pairwise (two $e$ charges or two $m$ charges) by means of single spin flips in the ground state. Their product is the fermionic excitation $em$, and these three charges, together with the vacuum charge $\mbb{I}$, exhibit the fusion rules and exchange statistics of the symmetry group $D(\mbb{Z}_2)$. More generally, the charges of an emergent topological order on a 2D manifold may exhibit fusion and braiding statistics derived from any unitary braided tensor category (though this is restricted to unitary \emph{modular} braided tensor categories if the manifold supports a non-trivial cycle). These generalised topologically ordered quasiparticle excitations are termed \emph{anyons},\cite{kitaev2006,bonderson2007,bonderson2008} 
and are of particular interest due to
their potential use in quantum computation\cite{nayak2008} and their proposed role in explaining the plateau states of the fractional quantum Hall effect.\cite{read1990,read1992,read1999,xia2004,pan2008,kumar2010} This interest is likely to be further enhanced by the recently reported detection of Ising anyons at the ends of iron nanowires.\cite{nadj-perge2014} Tensor networks provide one of the most promising approaches for the study of large-scale anyonic systems,\cite{pfeifer2010,konig2010} and therefore the application of accessory site iDMRG to anyonic systems is of particular importance.

When simulating any topologically ordered system using a tensor network algorithm, it is possible to protect the symmetries associated with the topological charges of the model. In the enforced absence of symmetry-breaking operations the topological charges behave as good quantum numbers, and may be exploited in tensor network algorithms for computational gain (and sometimes, in the absence of an explicit microscopic model, as a matter of necessity). The application of this approach to the finite DMRG algorithm is presented in \rcite{pfeifer2015}, and builds on the earlier work of \rcite{pfeifer2010} to develop a general toolbox for the construction and manipulation of anyonic tensors. This approach to DMRG for topologically ordered systems (including anyons) may be viewed as a generalisation of the exploitation of non-Abelian symmetries seen in earlier DMRG simulations.\cite{mcculloch2002}

While the toric code is most commonly studied (surprise, surprise) on the torus, 
other anyonic lattice systems are frequently studied on manifolds with a boundary, most commonly (and sometimes by implicit assumption) on the disc. However, any manifold boundary is topologically equivalent to one or more punctures, and thus for topologically ordered systems any
manifold with a boundary admits a (total) boundary charge. Much as enforcing a total spin sector restricts the admissible states in conventional DMRG, the choice of this charge %
may affect the family of states admissible on the lattice, even if no explicit Hamiltonian-mediated coupling between the lattice and the boundary exists.

For finite systems, as discussed in \rcite{pfeifer2015}, this boundary charge may be included as an additional site at one end of the chain, and this approach is directly analogous to the selection of a specific spin sector in conventional DMRG. The na\"\i{}eve generalisation of this approach to infinite systems is to include this charge on one side or the other of the initial state, as per \fref{fig:extendedv}, though we have already seen that this is suboptimal. The boundary charge may then be fixed by only admitting a single choice of charge on this boundary index, or multiple possible boundary charges may be admitted.

Once again, this treatment is improved upon by the approach described in \sref{sec:symIDMRG} where the boundary charge is expressed on the $\lambda$-tensor. As previously discussed, this results in a decrease in computational cost when the ground state may be expressed in block-periodic form \eref{eq:blockMPS}, and gives the ability to explicitly determine the boundary charge of the ground state by calculating weights as per \fref{fig:accMPS}(iii), and to restrict it by constraining the admissible charges on the accessory site. Finally, this approach delivers a further benefit when the boundary charge admits a real, physical interpretation: If the optimal boundary charge differs in the low-$L$ and high-$L$ regimes, then the inclusion of the accessory site in every update step facilitates the transition between these regimes, whereas if the boundary charge is included on the system or environment block as per \fref{fig:extendedv}, the qualitative behaviour of the low-$L$ regime may persist as charge boundary conditions are locked in by unfortunate choices of initial tensors $A^{(1)}$ and $B^{(1)}$.

\section{Further example\label{sec:example}}

As a more realistic example of the application of accessory site iDMRG, consider a system of hard-core $\mbb{Z}_3$ anyons with variable filling on a two-rung ladder. The only interactions are a term permitting anyons to hop to neighbouring sites if these are vacant, and a chemical potential. Braiding arises when two anyons use vacant spaces to hop around one another. Schematically, the Hamiltonian may be written
\begin{figure}
\includegraphics[width=\columnwidth]{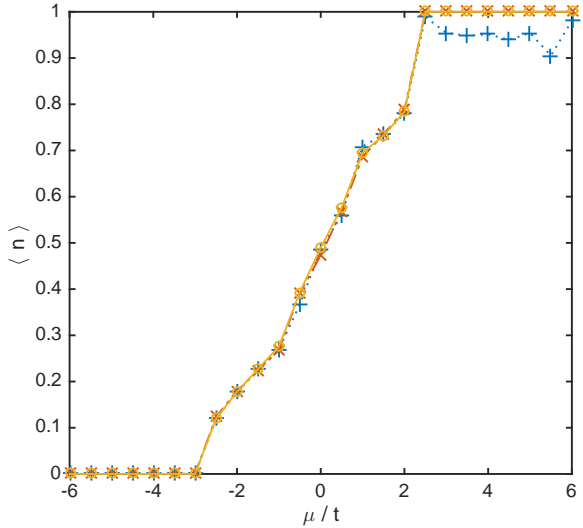}
\caption{Mean filling fraction for a system of free hard-core $\mbb{Z}_3$ anyons on a ladder \peref{eq:H}, after~300 iterations, as a function of $\mu/t$. Results are shown for conventional iDMRG ($+$), iDMRG with free boundary charges ($\times$), and accessory site iDMRG ($\circ$), all with MPS bond dimension $D=100$. 
Both iDMRG with free boundary charges and accessory site iDMRG correctly identify full occupation of the lattice %
at large values of $\mu/t$, but small discrepancies between these techniques are still observed at intermediate fillings. %
\label{fig:example}}
\end{figure}%
\begin{equation}
\hat H=-t\sum_{\la i,j\ra}\Big(\textrm{hopping terms}\Big)
-\mu\sum_i c^\dagger_i c_i\label{eq:H}
\end{equation}
where $\la i,j\ra$ represents all pairs of neighbouring sites.
Physical indices of the MPS may be in charge sector~0 or~1 corresponding to absence or presence of an anyon respectively, while MPS bond indices carry the cumulative anyonic charge, summing from left to right, which may be~0,~1, or~2. All tensors are symmetric, with nonzero entries only for combinations of charge labels whose oriented sum is zero. 
\Fref{fig:example} shows filling fraction as a function of $\mu/t$ after three hundred iterations.
Conventional iDMRG with exploitation of $\mbb{Z}_3$ symmetry is immediately seen to be unreliable at high filling fractions, for the reasons discussed in \sref{sec:illus}. Infinite DMRG with boundary charges and accessory site iDMRG both fare better at high filling fractions, correctly capturing the $\la n\ra=1$ insulating plateau, but show disagreement at intermediate fillings. 

\begin{figure}
\includegraphics[width=\linewidth]{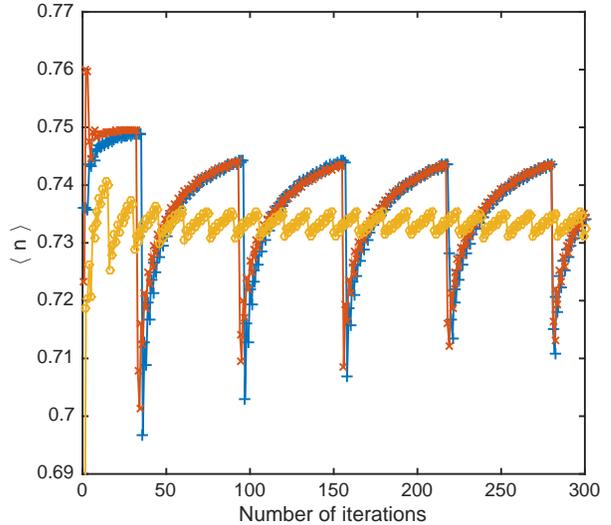}
\caption{Mean filling fraction $\la n\ra$ as a function of the number of iterations performed, for conventional iDMRG ($+$), iDMRG with free boundary charges ($\times$), and accessory site iDMRG ($\circ$), all at $\mu/t=1.5$ with MPS bond dimension $D=100$.\label{fig:convergence}}
\end{figure}

To understand this discrepancy, let $\mu/t$ be fixed at $1.5$ and consider how the value of $\la n \ra$ evolves as a function of the number of DMRG iterations performed, as shown in \fref{fig:convergence}.
At $\mu/t=1.5$ the filling fraction is sufficiently low that even the standard symmetric iDMRG algorithm does not suffer a catastrophic failure, yielding a superficially reasonable value for $\la n\ra$. Nevertheless, convergence of the standard algorithm is slow, exhibiting an extremely long periodicity of the sort described in Refs.~\olcite{caprara1997}--\olcite{caprara1997a}. The oscillations in $\la n\ra$ are still large after~300 iterations, and they exhibit noticeable variability from one cycle to the next, suggesting that the simulation is relatively poorly converged.

In contrast the periodicity for accessory site iDMRG is much shorter and the rate of convergence is faster, giving more accurate results after significantly fewer iterations%
. Comparison of the mean occupation number with a reference value obtained using a non-symmetric implementation of iDMRG is revealing: Protection of symmetry has forced the standard (symmetric) iDMRG algorithm to converge towards a much longer-period orbit than either the non-symmetric iDMRG algorithm or the accessory charge iDMRG algorithm (\fref{fig:convergence}), and the mean value of the filling fraction over this orbit differs from that obtained using the non-symmetric reference implementation (\tref{tab:results1}). 
\begin{table}[b]%
\caption{Mean filling fraction for Hamiltonian~\eref{eq:H} with $\mu/t=1.5$ after 300 iterations, for different variants of the iDMRG algorithm with MPS bond dimension $D=100$. The average $\la n\ra$ is taken over one orbit within the space of MPS states.\label{tab:results1}}
\begin{ruledtabular}
\begin{tabular}{l|c}
Algorithm & $\la n \ra$\\
\hline
Reference non-symmetric iDMRG & 0.73335 \\ %
Standard symmetric iDMRG & 0.73602 \\ %
Free boundary charges iDMRG & 0.73636 \\ %
Accessory site iDMRG & 0.73336 %
\end{tabular}
\end{ruledtabular}
\end{table}
This larger orbit must therefore be considered a symmetry-protected local minimum, while the non-symmetric iDMRG and accessory site iDMRG converge to orbits which more accurately approximate the true ground state.

Interestingly, at $\mu/t=1.5$ the free boundary charge algorithm is seen to settle into a similar longer-period orbit to the standard symmetric algorithm. For the free boundary charge algorithm this represents a relatively benign form of failure due to insufficiently large bond dimension and/or overconstrained boundary conditions at small chain lengths, as described in \sref{sec:extendedv}. Rather than recovering a translation-invariant representation of the ground state as per \Eref{eq:transinv}, the algorithm has broken translation symmetry in pursuit of lower energies. The benefit associated with the additional boundary charges has been lost, and the algorithm has ended up in a similar symmetry-protected local minimum to the standard symmetric iDMRG. Thus, while the free boundary charge algorithm is equally as effective as the accessory charge algorithm at sufficiently high and sufficiently low values of $\mu/t$, it is less reliable over an intermediate regime.

The generalisation of these observations to arbitrary Hamiltonians is as follows:
\begin{itemize}
\item Standard symmetric iDMRG: Avoids symmetry-protected local minima if the maximum achievable value of $\hat{\rho}^{(\mc{P})}$ in \Eref{eq:rhoP} is sufficiently small, as charge boundary conditions during initialisation of the MPS are not of critical importance.
\item Free boundary charge iDMRG: Avoids symmetry-protected local minima if the maximum achievable value of $\hat{\rho}^{(\mc{P})}$ is either sufficiently small, or sufficiently large and accompanied by a sufficiently large MPS bond dimension $D$. These conditions promote the retention of all significant charge sectors on the MPS bond. When the free boundary charge algorithm does become stuck in a local minimum, the effects are likely to be less severe as effective convergence may still be attained on the portions of the MPS tensors not participating in the intertwinors, when working in the intermediate $\hat{\rho}^{(\mc{P})}$ regime.
\item Accessory site iDMRG: Efficiently avoids symmetry-protected local minima for any maximum achievable value of $\hat{\rho}^{(\mc{P})}$ and any MPS bond dimension $D$.
\end{itemize}

\section{Implications for \lowercase{i}TEBD\label{sec:iTEBD}}

As explained in Sec.~III of \rcite{mcculloch2008}, there exists a deep connection between the iDMRG algorithm and the infinite Time-Evolving Block Decimation algorithm (iTEBD) with imaginary time evolution. Although these algorithms differ substantially in their usual presentation, they are in fact formally equivalent up to a choice of eigensolver used to find the ground state, with iDMRG employing the Lanczos method where iTEBD uses the power method, repeatedly applying an infinitesimal imaginary time evolution operator to a tensor pair then restoring normalisation of the wavefunction.

As a consequence, precisely the same local minima exist in symmetric iTEBD as have been seen here for iDMRG. The iTEBD algorithm, however, is much more robust against becoming stuck in a significant symmetry-protected local minimum. This is because the most common initialisation procedure for iTEBD is the selection of a random pair of initial tensors $A$ and $B$. The Ansatz for the infinite chain then comprises the pair $AB$ repeated an infinite number of times. The boundary conditions of this initialisation are equivalent to those given in \sref{sec:extendedv} for iDMRG, with all valid charges potentially being represented on the boundary. 
For sufficiently large MPS bond dimension $D$, the iTEBD algorithm will therefore converge towards a state having a form analogous to \Eref{eq:transinv}.

Nevertheless, as seen in \sref{sec:example}, it is still possible for iDMRG with free boundary charges---and therefore also for symmetric iTEBD---to become stuck in a symmetry-protected local minimum. However, this risk is only significant for intermediate values of $\hat{\rho}^{(\mc{P})}$, implying that only a part of each MPS tensor in the translation-invariant ground state exhibits an intertwinor structure. As seen in the example of \tref{tab:results1}, relatively good convergence to the true ground state is therefore possible, though the end precision is still less than that which can be obtained using accessory site iDMRG.

In summary, the standard iTEBD algorithm is therefore %
largely protected from the most damaging symmetry-protected local minima, but potentially performs suboptimally both in speed of convergence and in the precision obtainable for a given MPS bond dimension $D$. It is, however, unclear whether a counterpart to the accessory site iDMRG algorithm exists for iTEBD, as a necessary prerequisite for accessory site iDMRG is a numerical eigensolver which can switch between charge sectors from one iteration to the next. If this capability is suppressed, for example by using the initialisation vector estimation process of \aref{sec:initialvec}, then the accessory site algorithm is no better than standard iDMRG. 
At this time it is unclear how this ability to switch charge sectors might be incorporated into the iTEBD algorithm, and thus it is uncertain whether the iTEBD algorithm can 
be modified to 
obtain %
the enhanced performance seen in accessory site iDMRG.

\section{Conclusion}

Under certain circumstances, described in \sref{sec:intertwinors}, exploitation of symmetries in infinite DMRG may cause failure of simulations to converge to the correct ground state, becoming trapped instead in \emph{symmetry-protected local minima}. This paper has examined three modified forms of the infinite DMRG algorithm which permit reliable exploitation of symmetries of the Hamiltonian at high filling fraction. Of these, one (discussed in \sref{sec:blocksize} and \rcite{mcculloch2000}) requires \emph{a priori} knowledge of the repeated block length of the converged symmetric MPS and so is limited in its application. Another method (\sref{sec:extendedv}) is a generalisation of techniques used in finite DMRG to identify states having non-zero quantum number. This method performs reasonably well for an example realistic Hamiltonian, but may still encounter problems if the MPS bond dimension is insufficiently large, or if the boundary charges which are appropriate at initialisation are not well-suited to longer chains. %
The third method (\sref{sec:symIDMRG}) attaches an accessory site to the region undergoing variational update. %
This method successfully reproduces expected behaviour in all regimes, robustly re-evaluates the global boundary conditions on each iteration, offers improved rates of numerical convergence, %
and %
frequently requires a much smaller MPS bond dimension than the preceding method to obtain the same level of precision. When exploiting symmetries in iDMRG,
accessory site iDMRG %
is consequently the method of choice. %

\begin{figure}
\includegraphics[width=\columnwidth]{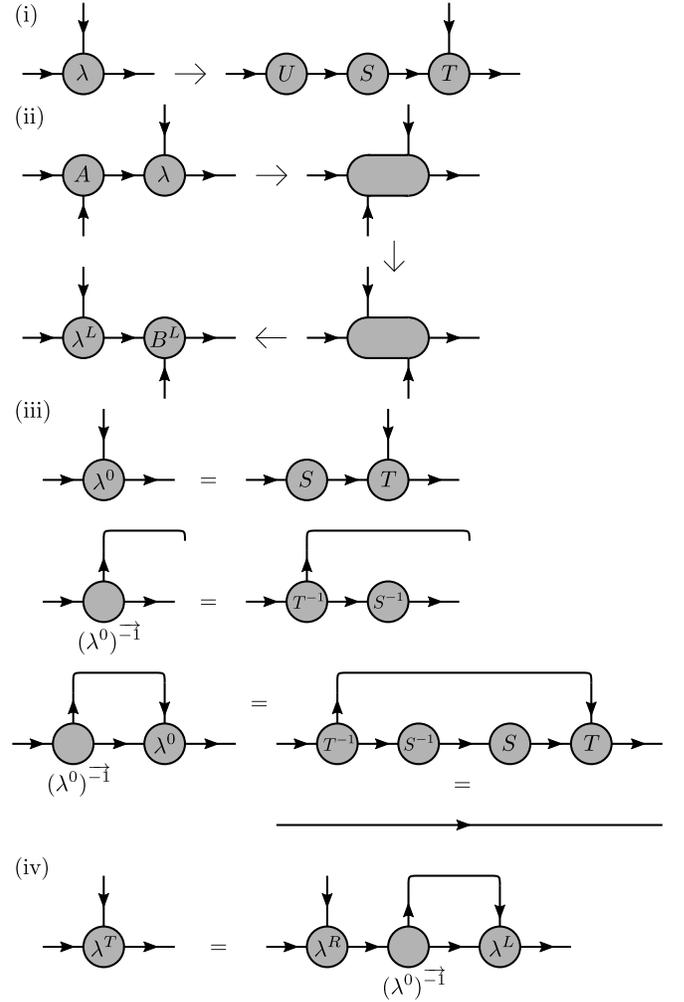}
\caption{(i)~Decomposition of $\lambda$ into a unitary matrix $U$, a diagonal matrix $S$ which carries the singular values, and an isometry $T$ which carries the accessory site. (ii)~Rolling tensor $\lambda$ left through tensor $A$. The decomposition is chosen such that $B^L$ is right-unitary as per \pfref{fig:iMPS}(i). (iii)~The left inverse of $\lambda^0$. (iv)~Construction of $\lambda^T$ using the left inverse of $\lambda^0$.\label{fig:rollupdate}}
\end{figure}%

From the relationship between iDMRG and iTEBD described in \rcite{mcculloch2008}, the boundary conditions in symmetric iTEBD are seen to be equivalent to those introduced in \sref{sec:extendedv}. A more comprehensive discussion may be found in Secs.~\ref{sec:example} and \ref{sec:iTEBD}, but in brief it follows that while iTEBD is relatively robust against symmetry-protected local minima, it may deliver results with precisions inferior to those of accessory site iDMRG for equivalent MPS bond dimension.

The accessory site algorithm also permits the study of systems where the extra site admits a physical interpretation, such as topologically ordered (e.g.~anyonic) systems on a manifold with a boundary, for which it represents the total boundary charge. In such systems the accessory charge may be left free (with its admissible value(s) in the ground state(s) being an output of the algorithm), or partially or fully constrained. The standard infinite DMRG algorithm (without an accessory site or boundary charges) is recovered if this site is constrained to hold the trivial charge.

\appendix
\section{Estimation of iDMRG update vector\label{sec:initialvec}}

This Appendix describes how the prescription given in \rcite{mcculloch2008} for estimating the eigenvector $v$ may be generalised when $v$ has an additional index corresponding to an accessory site. Its use is not recommended unless the charge on this accessory site is fixed by the system under study, as it is unable to accommodate cycling of the charge on the accessory site. Use of this prescription with the accessory site algorithm has been observed to cause simulations to converge to a %
state with fixed accessory charge when a lower-energy series of configurations with cyclic accessory charge (and the same MPS bond dimension) is known to exist. 

To generalise the prescription of \rcite{mcculloch2008}, first let $\lambda$ be decomposed as shown in \fref{fig:rollupdate}(i).
As a choice of gauge the unitary matrix may always be absorbed into the $A$-tensor to the immediate left, and thus $\lambda$ comprises a diagonal matrix $S$ containing singular values, and an isometry $T$ (a projector which obeys unitarity on the smaller Hilbert space) which carries the accessory site.

As in \rcite{mcculloch2008}, the process at the heart of the update procedure is ``rolling'' left and right, in which the location of the $\lambda$-tensor moves one bond to the left or the right on the MPS chain. To perform the process of ``rolling left'', for example, the $\lambda$-tensor and the immediately-adjacent $A$-tensor are fused, then separated again by means of singular value decomposition or $QR$ decomposition as shown in \fref{fig:rollupdate}(ii). (Details such as the sequence of fusing and splitting of indices, while necessary, are routine and so have not been explicitly specified.) The new $\lambda$-tensor, $\lambda^L$, is again gauged so that $U=\mbb{I}$ in \fref{fig:rollupdate}(i). Similarly, $\lambda^R$ is generated by rolling right. Finally, $\lambda^0$ denotes the optimal $\lambda$-tensor computed on the iteration before last.

It then suffices to construct either the left inverse or right inverse of $\lambda^0$, with the left inverse being shown in \fref{fig:rollupdate}(iii), and then constructing the trial vector
\begin{equation}
\lambda^T=\lambda^R\left(\lambda^0\right)^{-1}\lambda^L
\end{equation}
where the charge index of $\left(\lambda^0\right)^{-1}$ is summed with that of $\lambda^R$ if using the right inverse, or with that of $\lambda^L$ if using the left inverse [\fref{fig:rollupdate}(iv)]. The trial vector is then assembled in the usual way from $\lambda^T$ and the $A$- and $B$-tensors generated when rolling right and left (and incorporating any unitary matrices generated when gauging $\lambda^L$ and $\lambda^R$).


\begin{thebibliography}{47}%
\makeatletter
\providecommand \@ifxundefined [1]{%
 \@ifx{#1\undefined}
}%
\providecommand \@ifnum [1]{%
 \ifnum #1\expandafter \@firstoftwo
 \else \expandafter \@secondoftwo
 \fi
}%
\providecommand \@ifx [1]{%
 \ifx #1\expandafter \@firstoftwo
 \else \expandafter \@secondoftwo
 \fi
}%
\providecommand \natexlab [1]{#1}%
\providecommand \enquote  [1]{``#1''}%
\providecommand \bibnamefont  [1]{#1}%
\providecommand \bibfnamefont [1]{#1}%
\providecommand \citenamefont [1]{#1}%
\providecommand \href@noop [0]{\@secondoftwo}%
\providecommand \href [0]{\begingroup \@sanitize@url \@href}%
\providecommand \@href[1]{\@@startlink{#1}\@@href}%
\providecommand \@@href[1]{\endgroup#1\@@endlink}%
\providecommand \@sanitize@url [0]{\catcode `\\12\catcode `\$12\catcode
  `\&12\catcode `\#12\catcode `\^12\catcode `\_12\catcode `\%12\relax}%
\providecommand \@@startlink[1]{}%
\providecommand \@@endlink[0]{}%
\providecommand \url  [0]{\begingroup\@sanitize@url \@url }%
\providecommand \@url [1]{\endgroup\@href {#1}{\urlprefix }}%
\providecommand \urlprefix  [0]{URL }%
\providecommand \Eprint [0]{\href }%
\providecommand \doibase [0]{http://dx.doi.org/}%
\providecommand \selectlanguage [0]{\@gobble}%
\providecommand \bibinfo  [0]{\@secondoftwo}%
\providecommand \bibfield  [0]{\@secondoftwo}%
\providecommand \translation [1]{[#1]}%
\providecommand \BibitemOpen [0]{}%
\providecommand \bibitemStop [0]{}%
\providecommand \bibitemNoStop [0]{.\EOS\space}%
\providecommand \EOS [0]{\spacefactor3000\relax}%
\providecommand \BibitemShut  [1]{\csname bibitem#1\endcsname}%
\let\auto@bib@innerbib\@empty
\bibitem [{\citenamefont {White}(1992)}]{white1992}%
  \BibitemOpen
  \bibfield  {author} {\bibinfo {author} {\bibfnamefont {S.~R.}\ \bibnamefont
  {White}},\ }\href {\doibase 10.1103/PhysRevLett.69.2863} {\bibfield
  {journal} {\bibinfo  {journal} {Phys. Rev. Lett.}\ }\textbf {\bibinfo
  {volume} {69}},\ \bibinfo {pages} {2863} (\bibinfo {year}
  {1992})}\BibitemShut {NoStop}%
\bibitem [{\citenamefont {White}(1993)}]{white1993}%
  \BibitemOpen
  \bibfield  {author} {\bibinfo {author} {\bibfnamefont {S.~R.}\ \bibnamefont
  {White}},\ }\href {\doibase 10.1103/PhysRevB.48.10345} {\bibfield  {journal}
  {\bibinfo  {journal} {Phys. Rev. B}\ }\textbf {\bibinfo {volume} {48}},\
  \bibinfo {pages} {10345} (\bibinfo {year} {1993})}\BibitemShut {NoStop}%
\bibitem [{\citenamefont {Schollw\"{o}ck}(2005)}]{schollwock2005}%
  \BibitemOpen
  \bibfield  {author} {\bibinfo {author} {\bibfnamefont {U.}~\bibnamefont
  {Schollw\"{o}ck}},\ }\href {\doibase 10.1103/RevModPhys.77.259} {\bibfield
  {journal} {\bibinfo  {journal} {Rev. Mod. Phys.}\ }\textbf {\bibinfo {volume}
  {77}},\ \bibinfo {eid} {259} (\bibinfo {year} {2005})}\BibitemShut {NoStop}%
\bibitem [{\citenamefont {Schollw\"{o}ck}(2011)}]{schollwock2011}%
  \BibitemOpen
  \bibfield  {author} {\bibinfo {author} {\bibfnamefont {U.}~\bibnamefont
  {Schollw\"{o}ck}},\ }\href {\doibase 10.1016/j.aop.2010.09.012} {\bibfield
  {journal} {\bibinfo  {journal} {Ann. Phys.}\ }\textbf {\bibinfo {volume}
  {326}},\ \bibinfo {pages} {96} (\bibinfo {year} {2011})}\BibitemShut
  {NoStop}%
\bibitem [{\citenamefont {\"Ostlund}\ and\ \citenamefont
  {Rommer}(1995)}]{ostlund1995}%
  \BibitemOpen
  \bibfield  {author} {\bibinfo {author} {\bibfnamefont {S.}~\bibnamefont
  {\"Ostlund}}\ and\ \bibinfo {author} {\bibfnamefont {S.}~\bibnamefont
  {Rommer}},\ }\href {\doibase 10.1103/PhysRevLett.75.3537} {\bibfield
  {journal} {\bibinfo  {journal} {Phys. Rev. Lett.}\ }\textbf {\bibinfo
  {volume} {75}},\ \bibinfo {pages} {3537} (\bibinfo {year}
  {1995})}\BibitemShut {NoStop}%
\bibitem [{\citenamefont {Verstraete}\ \emph {et~al.}(2004)\citenamefont
  {Verstraete}, \citenamefont {Porras},\ and\ \citenamefont
  {Cirac}}]{verstraete2004}%
  \BibitemOpen
  \bibfield  {author} {\bibinfo {author} {\bibfnamefont {F.}~\bibnamefont
  {Verstraete}}, \bibinfo {author} {\bibfnamefont {D.}~\bibnamefont {Porras}},
  \ and\ \bibinfo {author} {\bibfnamefont {J.~I.}\ \bibnamefont {Cirac}},\
  }\href {\doibase 10.1103/PhysRevLett.93.227205} {\bibfield  {journal}
  {\bibinfo  {journal} {Phys. Rev. Lett.}\ }\textbf {\bibinfo {volume} {93}},\
  \bibinfo {pages} {227205} (\bibinfo {year} {2004})}\BibitemShut {NoStop}%
\bibitem [{\citenamefont {McCulloch}(2007)}]{mcculloch2007}%
  \BibitemOpen
  \bibfield  {author} {\bibinfo {author} {\bibfnamefont {I.~P.}\ \bibnamefont
  {McCulloch}},\ }\href {http://stacks.iop.org/1742-5468/2007/i=10/a=P10014}
  {\bibfield  {journal} {\bibinfo  {journal} {J. Stat. Mech.: Theory and Exp.}\
  }\textbf {\bibinfo {volume} {2007}},}\ \href {http://stacks.iop.org/1742-5468/2007/i=10/a=P10014}{\bibinfo {pages} {P10014} (\bibinfo
  {year} {2007})}\BibitemShut {NoStop}%
\bibitem [{\citenamefont {McCulloch}\ and\ \citenamefont
  {Gul\'acsi}(2000)}]{mcculloch2000}%
  \BibitemOpen
  \bibfield  {author} {\bibinfo {author} {\bibfnamefont {I.~P.}\ \bibnamefont
  {McCulloch}}\ and\ \bibinfo {author} {\bibfnamefont {M.}~\bibnamefont
  {Gul\'acsi}},\ }\href {http://www.publish.csiro.au/paper/PH00023} {\bibfield
  {journal} {\bibinfo  {journal} {Australian J. Phys.}\ }\textbf {\bibinfo
  {volume} {53}},\ \bibinfo {pages} {597} (\bibinfo {year} {2000})}\BibitemShut
  {NoStop}%
\bibitem [{\citenamefont {McCulloch}\ and\ \citenamefont
  {Gul{\'a}csi}(2002)}]{mcculloch2002}%
  \BibitemOpen
  \bibfield  {author} {\bibinfo {author} {\bibfnamefont {I.~P.}\ \bibnamefont
  {McCulloch}}\ and\ \bibinfo {author} {\bibfnamefont {M.}~\bibnamefont
  {Gul{\'a}csi}},\ }\href {http://stacks.iop.org/0295-5075/57/i=6/a=852}
  {\bibfield  {journal} {\bibinfo  {journal} {Europhys. Lett.}\ }\textbf
  {\bibinfo {volume} {57}},\ \bibinfo {pages} {852} (\bibinfo {year}
  {2002})}\BibitemShut {NoStop}%
\bibitem [{\citenamefont {Singh}\ \emph {et~al.}(2010)\citenamefont {Singh},
  \citenamefont {Pfeifer},\ and\ \citenamefont {Vidal}}]{singh2010}%
  \BibitemOpen
  \bibfield  {author} {\bibinfo {author} {\bibfnamefont {S.}~\bibnamefont
  {Singh}}, \bibinfo {author} {\bibfnamefont {R.~N.~C.}\ \bibnamefont
  {Pfeifer}}, \ and\ \bibinfo {author} {\bibfnamefont {G.}~\bibnamefont
  {Vidal}},\ }\href {\doibase 10.1103/PhysRevA.82.050301} {\bibfield  {journal}
  {\bibinfo  {journal} {Phys. Rev. A}\ }\textbf {\bibinfo {volume} {82}},\
  \bibinfo {pages} {050301} (\bibinfo {year} {2010})}\BibitemShut {NoStop}%
\bibitem [{\citenamefont {Singh}\ \emph {et~al.}(2011)\citenamefont {Singh},
  \citenamefont {Pfeifer},\ and\ \citenamefont {Vidal}}]{singh2011}%
  \BibitemOpen
  \bibfield  {author} {\bibinfo {author} {\bibfnamefont {S.}~\bibnamefont
  {Singh}}, \bibinfo {author} {\bibfnamefont {R.~N.~C.}\ \bibnamefont
  {Pfeifer}}, \ and\ \bibinfo {author} {\bibfnamefont {G.}~\bibnamefont
  {Vidal}},\ }\href {\doibase 10.1103/PhysRevB.83.115125} {\bibfield  {journal}
  {\bibinfo  {journal} {Phys. Rev. B}\ }\textbf {\bibinfo {volume} {83}},\
  \bibinfo {pages} {115125} (\bibinfo {year} {2011})}\BibitemShut {NoStop}%
\bibitem [{\citenamefont {Pfeifer}(2011)}]{pfeifer2011a}%
  \BibitemOpen
  \bibfield  {author} {\bibinfo {author} {\bibfnamefont {R.~N.~C.}\
  \bibnamefont {Pfeifer}},\ }\emph {\bibinfo {title} {Simulation of Anyons
  Using Symmetric Tensor Network Algorithms}},\ \href@noop {} {Ph.D. thesis},\
  \bibinfo  {school} {The University of Queensland} (\bibinfo {year} {2011}),\
  \Eprint {http://arxiv.org/abs/1202.1522v2}
  {arXiv:1202.1522v2 [cond-mat.str-el]} \BibitemShut {NoStop}%
\bibitem [{\citenamefont {Singh}\ and\ \citenamefont
  {Vidal}(2012)}]{singh2012}%
  \BibitemOpen
  \bibfield  {author} {\bibinfo {author} {\bibfnamefont {S.}~\bibnamefont
  {Singh}}\ and\ \bibinfo {author} {\bibfnamefont {G.}~\bibnamefont {Vidal}},\
  }\href {\doibase 10.1103/PhysRevB.86.195114} {\bibfield  {journal} {\bibinfo
  {journal} {Phys. Rev. B}\ }\textbf {\bibinfo {volume} {86}},\ \bibinfo
  {pages} {195114} (\bibinfo {year} {2012})}\BibitemShut {NoStop}%
\bibitem [{\citenamefont {Caprara}\ and\ \citenamefont
  {Rosengren}(1997{\natexlab{a}})}]{caprara1997}%
  \BibitemOpen
  \bibfield  {author} {\bibinfo {author} {\bibfnamefont {S.}~\bibnamefont
  {Caprara}}\ and\ \bibinfo {author} {\bibfnamefont {A.}~\bibnamefont
  {Rosengren}},\ }\href {http://stacks.iop.org/0295-5075/39/i=1/a=055}
  {\bibfield  {journal} {\bibinfo  {journal} {Europhys. Lett.}\ }\textbf
  {\bibinfo {volume} {39}},\ \bibinfo {pages} {55} (\bibinfo {year}
  {1997}{\natexlab{a}})}\BibitemShut {NoStop}%
\bibitem [{\citenamefont {Caprara}\ and\ \citenamefont
  {Rosengren}(1997{\natexlab{b}})}]{caprara1997a}%
  \BibitemOpen
  \bibfield  {author} {\bibinfo {author} {\bibfnamefont {S.}~\bibnamefont
  {Caprara}}\ and\ \bibinfo {author} {\bibfnamefont {A.}~\bibnamefont
  {Rosengren}},\ }\href {\doibase
  http://dx.doi.org/10.1016/S0550-3213(97)00142-9} {\bibfield  {journal}
  {\bibinfo  {journal} {Nuc. Phys. B}\ }\textbf {\bibinfo {volume} {493}},\
  \bibinfo {pages} {640 } (\bibinfo {year} {1997}{\natexlab{b}})}\BibitemShut
  {NoStop}%
\bibitem [{\citenamefont {Kitaev}(2006)}]{kitaev2006}%
  \BibitemOpen
  \bibfield  {author} {\bibinfo {author} {\bibfnamefont {A.}~\bibnamefont
  {Kitaev}},\ }\href {\doibase DOI: 10.1016/j.aop.2005.10.005} {\bibfield
  {journal} {\bibinfo  {journal} {Ann. Phys.}\ }\textbf {\bibinfo {volume}
  {321}},\ \bibinfo {pages} {2 } (\bibinfo {year} {2006})}\BibitemShut
  {NoStop}%
\bibitem [{\citenamefont {Bonderson}(2007)}]{bonderson2007}%
  \BibitemOpen
  \bibfield  {author} {\bibinfo {author} {\bibfnamefont {P.~H.}\ \bibnamefont
  {Bonderson}},\ }\emph {\bibinfo {title} {Non-Abelian Anyons and
  Interferometry}},\ \href
  {http://resolver.caltech.edu/CaltechETD:etd-06042007-101617} {Ph.D. thesis},\
  \bibinfo  {school} {California Institute of Technology} (\bibinfo {year}
  {2007})\BibitemShut {NoStop}%
\bibitem [{\citenamefont {Bonderson}\ \emph {et~al.}(2008)\citenamefont
  {Bonderson}, \citenamefont {Shtengel},\ and\ \citenamefont
  {Slingerland}}]{bonderson2008}%
  \BibitemOpen
  \bibfield  {author} {\bibinfo {author} {\bibfnamefont {P.}~\bibnamefont
  {Bonderson}}, \bibinfo {author} {\bibfnamefont {K.}~\bibnamefont {Shtengel}},
  \ and\ \bibinfo {author} {\bibfnamefont {J.}~\bibnamefont {Slingerland}},\
  }\href {\doibase DOI: 10.1016/j.aop.2008.01.012} {\bibfield  {journal}
  {\bibinfo  {journal} {Ann. Phys.}\ }\textbf {\bibinfo {volume} {323}},\
  \bibinfo {pages} {2709 } (\bibinfo {year} {2008})}\BibitemShut {NoStop}%
\bibitem [{\citenamefont {Pfeifer}\ \emph {et~al.}(2012)\citenamefont
  {Pfeifer}, \citenamefont {Buerschaper}, \citenamefont {Trebst}, \citenamefont
  {Ludwig}, \citenamefont {Troyer},\ and\ \citenamefont
  {Vidal}}]{pfeifer2012a}%
  \BibitemOpen
  \bibfield  {author} {\bibinfo {author} {\bibfnamefont {R.~N.~C.}\
  \bibnamefont {Pfeifer}}, \bibinfo {author} {\bibfnamefont {O.}~\bibnamefont
  {Buerschaper}}, \bibinfo {author} {\bibfnamefont {S.}~\bibnamefont {Trebst}},
  \bibinfo {author} {\bibfnamefont {A.~W.~W.}\ \bibnamefont {Ludwig}}, \bibinfo
  {author} {\bibfnamefont {M.}~\bibnamefont {Troyer}}, \ and\ \bibinfo {author}
  {\bibfnamefont {G.}~\bibnamefont {Vidal}},\ }\href {\doibase
  10.1103/PhysRevB.86.155111} {\bibfield  {journal} {\bibinfo  {journal} {Phys.
  Rev. B}\ }\textbf {\bibinfo {volume} {86}},\ \bibinfo {pages} {155111}
  (\bibinfo {year} {2012})}\BibitemShut {NoStop}%
\bibitem [{Note1()}]{Note1}%
  \BibitemOpen
  \bibinfo {note} {For \protect {$\varepsilon =0$} and infinite MPS bond
  dimension \protect {$D$} it is in theory possible to construct a Hamiltonian
  for which \protect {$p$} is infinite; an example would be a system whose
  filling fraction tends towards an irrational number as \protect
  {$k\rightarrow \infty $}, with particle number a protected symmetry. By
  definition such a system does not converge for any finite $k$. In practice,
  with finite \protect {$D$} and \protect {$\varepsilon $} and a deterministic
  computing device, the state space of the calculation is finite and any
  trajectory through this space will eventually repeat for some \protect {$p$},
  which may or may not be small enough to be recognised in
  practice.}\BibitemShut {Stop}%
\bibitem [{\citenamefont {Pfeifer}()}]{pfeifer2015a}%
  \BibitemOpen
  \bibfield  {author} {\bibinfo {author} {\bibfnamefont {R.~N.~C.}\
  \bibnamefont {Pfeifer}},\ }\href {http://arxiv.org/abs/1505.06928} {}\Eprint
  {http://arxiv.org/abs/1505.06928}
  {arXiv:1505.06928 [cond-mat.str-el] (2015)} \BibitemShut {NoStop}%
\bibitem [{Note2()}]{Note2}%
  \BibitemOpen
  \bibinfo {note} {The notion of locality on the space of matrix product states
  is not, in general, rigorously defined. A state \protect {$|\psi '\delimiter
  "526930B $} may be thought of as local with respect to a state \protect
  {$|\psi \delimiter "526930B $}, a Hamiltonian \protect {$\protect
  \mathaccentV {hat}05E{H}$}, and an algorithm \protect {$\protect \mathcal
  {A}$} if a single iteration of algorithm \protect {$\protect \mathcal {A}$}
  with Hamiltonian \protect {$\protect \mathaccentV {hat}05E{H}$} is capable of
  taking \protect {$|\psi \delimiter "526930B $} into \protect {$|\psi
  '\delimiter "526930B $}.}\BibitemShut {Stop}%
\bibitem [{Note3()}]{Note3}%
  \BibitemOpen
  \bibinfo {note} {If \protect {$\protect \mathaccentV {hat}05EH$} is invariant
  under translation by \protect {$p$} sites, a translation-invariant
  representation of \protect {$\protect \mathaccentV {hat}05EH$} may be
  constructed by fusing \protect {$p$} sites of dimension \protect {$d$} into
  one effective site of dimension \protect {$d^p$}.}\BibitemShut {Stop}%
\bibitem [{\citenamefont {McCulloch}\ \emph {et~al.}(1999)\citenamefont
  {McCulloch}, \citenamefont {Gulacsi}, \citenamefont {Caprara}, \citenamefont
  {Jazavaou},\ and\ \citenamefont {Rosengren}}]{mcculloch1999}%
  \BibitemOpen
  \bibfield  {author} {\bibinfo {author} {\bibfnamefont {I.}~\bibnamefont
  {McCulloch}}, \bibinfo {author} {\bibfnamefont {M.}~\bibnamefont {Gulacsi}},
  \bibinfo {author} {\bibfnamefont {S.}~\bibnamefont {Caprara}}, \bibinfo
  {author} {\bibfnamefont {A.}~\bibnamefont {Jazavaou}}, \ and\ \bibinfo
  {author} {\bibfnamefont {A.}~\bibnamefont {Rosengren}},\ }\href {\doibase
  10.1023/A:1022557314114} {\bibfield  {journal} {\bibinfo  {journal} {J. Low
  Temp. Phys.}\ }\textbf {\bibinfo {volume} {117}},\ \bibinfo {pages} {323}
  (\bibinfo {year} {1999})}\BibitemShut {NoStop}%
\bibitem [{\citenamefont {Vidal}(2004)}]{vidal2004}%
  \BibitemOpen
  \bibfield  {author} {\bibinfo {author} {\bibfnamefont {G.}~\bibnamefont
  {Vidal}},\ }\href {\doibase 10.1103/PhysRevLett.93.040502} {\bibfield
  {journal} {\bibinfo  {journal} {Phys. Rev. Lett.}\ }\textbf {\bibinfo
  {volume} {93}},\ \bibinfo {pages} {040502} (\bibinfo {year}
  {2004})}\BibitemShut {NoStop}%
\bibitem [{Note4()}]{Note4}%
  \BibitemOpen
  \bibinfo {note} {Note that the algorithm as a whole may still display
  periodicity---it is only periodicity of charge configurations which is
  suppressed and which causes increased computational cost. For example, in
  Refs.~\protect {\protect \rev@citealpnum {caprara1997}}--\protect {\protect
  \rev@citealpnum {caprara1997a}} conservation of particle number is not
  exploited for computational gain. The results are asymptotically \protect
  {$p$}-periodic with \protect {$p>1$}, but all indices carry only the trivial
  charge.}\BibitemShut {Stop}%
\bibitem [{Note5()}]{Note5}%
  \BibitemOpen
  \bibinfo {note} {A superposition of charge sectors may be selected if both
  yield equally low energy states, and \protect {$D$} is redundantly
  large.}\BibitemShut {Stop}%
\bibitem [{\citenamefont {McCulloch}()}]{mcculloch2008}%
  \BibitemOpen
  \bibfield  {author} {\bibinfo {author} {\bibfnamefont {I.~P.}\ \bibnamefont
  {McCulloch}},\ }\href {http://arxiv.org/abs/0804.2509v1} {}\Eprint
  {http://arxiv.org/abs/0804.2509v1}
  {arXiv:0804.2509v1 [cond-mat.str-el] (2008)} \BibitemShut {NoStop}%
\bibitem [{\citenamefont {Nishino}\ and\ \citenamefont
  {Okunishi}(1995)}]{nishino1995}%
  \BibitemOpen
  \bibfield  {author} {\bibinfo {author} {\bibfnamefont {T.}~\bibnamefont
  {Nishino}}\ and\ \bibinfo {author} {\bibfnamefont {K.}~\bibnamefont
  {Okunishi}},\ }\href {\doibase 10.1143/JPSJ.64.4084} {\bibfield  {journal}
  {\bibinfo  {journal} {J. Phys. Soc. Jpn.}\ }\textbf {\bibinfo {volume}
  {64}},\ \bibinfo {pages} {4084} (\bibinfo {year} {1995})}\BibitemShut
  {NoStop}%
\bibitem [{\citenamefont {Hieida}\ \emph {et~al.}(1997)\citenamefont {Hieida},
  \citenamefont {Okunishi},\ and\ \citenamefont {Akutsu}}]{hieida1997}%
  \BibitemOpen
  \bibfield  {author} {\bibinfo {author} {\bibfnamefont {Y.}~\bibnamefont
  {Hieida}}, \bibinfo {author} {\bibfnamefont {K.}~\bibnamefont {Okunishi}}, \
  and\ \bibinfo {author} {\bibfnamefont {Y.}~\bibnamefont {Akutsu}},\ }\href
  {\doibase http://dx.doi.org/10.1016/S0375-9601(97)00498-2} {\bibfield
  {journal} {\bibinfo  {journal} {Phys. Lett. A}\ }\textbf {\bibinfo {volume}
  {233}},\ \bibinfo {pages} {464 } (\bibinfo {year} {1997})}\BibitemShut
  {NoStop}%
\bibitem [{\citenamefont {Okunishi}\ \emph {et~al.}(1999)\citenamefont
  {Okunishi}, \citenamefont {Hieida},\ and\ \citenamefont
  {Akutsu}}]{okunishi1999}%
  \BibitemOpen
  \bibfield  {author} {\bibinfo {author} {\bibfnamefont {K.}~\bibnamefont
  {Okunishi}}, \bibinfo {author} {\bibfnamefont {Y.}~\bibnamefont {Hieida}}, \
  and\ \bibinfo {author} {\bibfnamefont {Y.}~\bibnamefont {Akutsu}},\ }\href
  {\doibase 10.1103/PhysRevB.59.6806} {\bibfield  {journal} {\bibinfo
  {journal} {Phys. Rev. B}\ }\textbf {\bibinfo {volume} {59}},\ \bibinfo
  {pages} {6806} (\bibinfo {year} {1999})}\BibitemShut {NoStop}%
\bibitem [{\citenamefont {Ueda}\ \emph {et~al.}(2006)\citenamefont {Ueda},
  \citenamefont {Nishino}, \citenamefont {Okunishi}, \citenamefont {Hieida},
  \citenamefont {Derian},\ and\ \citenamefont {Gendiar}}]{ueda2006}%
  \BibitemOpen
  \bibfield  {author} {\bibinfo {author} {\bibfnamefont {K.}~\bibnamefont
  {Ueda}}, \bibinfo {author} {\bibfnamefont {T.}~\bibnamefont {Nishino}},
  \bibinfo {author} {\bibfnamefont {K.}~\bibnamefont {Okunishi}}, \bibinfo
  {author} {\bibfnamefont {Y.}~\bibnamefont {Hieida}}, \bibinfo {author}
  {\bibfnamefont {R.}~\bibnamefont {Derian}}, \ and\ \bibinfo {author}
  {\bibfnamefont {A.}~\bibnamefont {Gendiar}},\ }\href {\doibase
  10.1143/JPSJ.75.014003} {\bibfield  {journal} {\bibinfo  {journal} {J. Phys.
  Soc. Jpn.}\ }\textbf {\bibinfo {volume} {75}},\ \bibinfo {pages} {014003}
  (\bibinfo {year} {2006})}\BibitemShut {NoStop}%
\bibitem [{\citenamefont {Kitaev}(2003)}]{kitaev2003}%
  \BibitemOpen
  \bibfield  {author} {\bibinfo {author} {\bibfnamefont {A.~Y.}\ \bibnamefont
  {Kitaev}},\ }\href {\doibase 10.1016/S0003-4916(02)00018-0} {\bibfield
  {journal} {\bibinfo  {journal} {Ann. Phys.}\ }\textbf {\bibinfo {volume}
  {303}},\ \bibinfo {pages} {2} (\bibinfo {year} {2003})}\BibitemShut {NoStop}%
\bibitem [{\citenamefont {{Di Francesco}}\ \emph {et~al.}(1997)\citenamefont
  {{Di Francesco}}, \citenamefont {Mathieu},\ and\ \citenamefont
  {S\'{e}n\'{e}chal}}]{difrancesco1997}%
  \BibitemOpen
  \bibfield  {author} {\bibinfo {author} {\bibfnamefont {P.}~\bibnamefont {{Di
  Francesco}}}, \bibinfo {author} {\bibfnamefont {P.}~\bibnamefont {Mathieu}},
  \ and\ \bibinfo {author} {\bibfnamefont {D.}~\bibnamefont
  {S\'{e}n\'{e}chal}},\ }\href@noop {} {\emph {\bibinfo {title} {Conformal
  Field Theory}}}\ (\bibinfo  {publisher} {Springer},\ \bibinfo {address} {New
  York},\ \bibinfo {year} {1997})\BibitemShut {NoStop}%
\bibitem [{\citenamefont {Pfeifer}\ \emph {et~al.}(2009)\citenamefont
  {Pfeifer}, \citenamefont {Evenbly},\ and\ \citenamefont
  {Vidal}}]{pfeifer2009}%
  \BibitemOpen
  \bibfield  {author} {\bibinfo {author} {\bibfnamefont {R.~N.~C.}\
  \bibnamefont {Pfeifer}}, \bibinfo {author} {\bibfnamefont {G.}~\bibnamefont
  {Evenbly}}, \ and\ \bibinfo {author} {\bibfnamefont {G.}~\bibnamefont
  {Vidal}},\ }\href {\doibase 10.1103/PhysRevA.79.040301} {\bibfield  {journal}
  {\bibinfo  {journal} {Phys. Rev. A}\ }\textbf {\bibinfo {volume} {79}},\
  \bibinfo {pages} {040301} (\bibinfo {year} {2009})}\BibitemShut {NoStop}%
\bibitem [{\citenamefont {Pfeifer}\ \emph {et~al.}(2010)\citenamefont
  {Pfeifer}, \citenamefont {Corboz}, \citenamefont {Buerschaper}, \citenamefont
  {Aguado}, \citenamefont {Troyer},\ and\ \citenamefont {Vidal}}]{pfeifer2010}%
  \BibitemOpen
  \bibfield  {author} {\bibinfo {author} {\bibfnamefont {R.~N.~C.}\
  \bibnamefont {Pfeifer}}, \bibinfo {author} {\bibfnamefont {P.}~\bibnamefont
  {Corboz}}, \bibinfo {author} {\bibfnamefont {O.}~\bibnamefont {Buerschaper}},
  \bibinfo {author} {\bibfnamefont {M.}~\bibnamefont {Aguado}}, \bibinfo
  {author} {\bibfnamefont {M.}~\bibnamefont {Troyer}}, \ and\ \bibinfo {author}
  {\bibfnamefont {G.}~\bibnamefont {Vidal}},\ }\href {\doibase
  10.1103/PhysRevB.82.115126} {\bibfield  {journal} {\bibinfo  {journal} {Phys.
  Rev. B}\ }\textbf {\bibinfo {volume} {82}},\ \bibinfo {pages} {115126}
  (\bibinfo {year} {2010})}\BibitemShut {NoStop}%
\bibitem [{\citenamefont {K\"onig}\ and\ \citenamefont
  {Bilgin}(2010)}]{konig2010}%
  \BibitemOpen
  \bibfield  {author} {\bibinfo {author} {\bibfnamefont {R.}~\bibnamefont
  {K\"onig}}\ and\ \bibinfo {author} {\bibfnamefont {E.}~\bibnamefont
  {Bilgin}},\ }\href {\doibase 10.1103/PhysRevB.82.125118} {\bibfield
  {journal} {\bibinfo  {journal} {Phys. Rev. B}\ }\textbf {\bibinfo {volume}
  {82}},\ \bibinfo {pages} {125118} (\bibinfo {year} {2010})}\BibitemShut
  {NoStop}%
\bibitem [{\citenamefont {Pfeifer}(2012)}]{pfeifer2012}%
  \BibitemOpen
  \bibfield  {author} {\bibinfo {author} {\bibfnamefont {R.~N.~C.}\
  \bibnamefont {Pfeifer}},\ }\href {\doibase 10.1103/PhysRevB.85.245126}
  {\bibfield  {journal} {\bibinfo  {journal} {Phys. Rev. B}\ }\textbf {\bibinfo
  {volume} {85}},\ \bibinfo {pages} {245126} (\bibinfo {year}
  {2012})}\BibitemShut {NoStop}%
\bibitem [{\citenamefont {Nayak}\ \emph {et~al.}(2008)\citenamefont {Nayak},
  \citenamefont {Simon}, \citenamefont {Stern}, \citenamefont {Freedman},\ and\
  \citenamefont {Das~Sarma}}]{nayak2008}%
  \BibitemOpen
  \bibfield  {author} {\bibinfo {author} {\bibfnamefont {C.}~\bibnamefont
  {Nayak}}, \bibinfo {author} {\bibfnamefont {S.~H.}\ \bibnamefont {Simon}},
  \bibinfo {author} {\bibfnamefont {A.}~\bibnamefont {Stern}}, \bibinfo
  {author} {\bibfnamefont {M.}~\bibnamefont {Freedman}}, \ and\ \bibinfo
  {author} {\bibfnamefont {S.}~\bibnamefont {Das~Sarma}},\ }\href {\doibase
  10.1103/RevModPhys.80.1083} {\bibfield  {journal} {\bibinfo  {journal} {Rev.
  Mod. Phys.}\ }\textbf {\bibinfo {volume} {80}},\ \bibinfo {pages} {1083}
  (\bibinfo {year} {2008})}\BibitemShut {NoStop}%
\bibitem [{\citenamefont {Read}(1990)}]{read1990}%
  \BibitemOpen
  \bibfield  {author} {\bibinfo {author} {\bibfnamefont {N.}~\bibnamefont
  {Read}},\ }\href {\doibase 10.1103/PhysRevLett.65.1502} {\bibfield  {journal}
  {\bibinfo  {journal} {Phys. Rev. Lett.}\ }\textbf {\bibinfo {volume} {65}},\
  \bibinfo {pages} {1502} (\bibinfo {year} {1990})}\BibitemShut {NoStop}%
\bibitem [{\citenamefont {Read}\ and\ \citenamefont {Moore}(1992)}]{read1992}%
  \BibitemOpen
  \bibfield  {author} {\bibinfo {author} {\bibfnamefont {N.}~\bibnamefont
  {Read}}\ and\ \bibinfo {author} {\bibfnamefont {G.}~\bibnamefont {Moore}},\
  }\href {\doibase 10.1143/PTPS.107.157} {\bibfield  {journal} {\bibinfo
  {journal} {Prog. Theo. Phys. Supp.}\ }\textbf {\bibinfo {volume} {107}},\
  \bibinfo {pages} {157} (\bibinfo {year} {1992})}\BibitemShut {NoStop}%
\bibitem [{\citenamefont {Read}\ and\ \citenamefont {Rezayi}(1999)}]{read1999}%
  \BibitemOpen
  \bibfield  {author} {\bibinfo {author} {\bibfnamefont {N.}~\bibnamefont
  {Read}}\ and\ \bibinfo {author} {\bibfnamefont {E.}~\bibnamefont {Rezayi}},\
  }\href {\doibase 10.1103/PhysRevB.59.8084} {\bibfield  {journal} {\bibinfo
  {journal} {Phys. Rev. B}\ }\textbf {\bibinfo {volume} {59}},\ \bibinfo
  {pages} {8084} (\bibinfo {year} {1999})}\BibitemShut {NoStop}%
\bibitem [{\citenamefont {Xia}\ \emph {et~al.}(2004)\citenamefont {Xia},
  \citenamefont {Pan}, \citenamefont {Vicente}, \citenamefont {Adams},
  \citenamefont {Sullivan}, \citenamefont {Stormer}, \citenamefont {Tsui},
  \citenamefont {Pfeiffer}, \citenamefont {Baldwin},\ and\ \citenamefont
  {West}}]{xia2004}%
  \BibitemOpen
  \bibfield  {author} {\bibinfo {author} {\bibfnamefont {J.~S.}\ \bibnamefont
  {Xia}}, \bibinfo {author} {\bibfnamefont {W.}~\bibnamefont {Pan}}, \bibinfo
  {author} {\bibfnamefont {C.~L.}\ \bibnamefont {Vicente}}, \bibinfo {author}
  {\bibfnamefont {E.~D.}\ \bibnamefont {Adams}}, \bibinfo {author}
  {\bibfnamefont {N.~S.}\ \bibnamefont {Sullivan}}, \bibinfo {author}
  {\bibfnamefont {H.~L.}\ \bibnamefont {Stormer}}, \bibinfo {author}
  {\bibfnamefont {D.~C.}\ \bibnamefont {Tsui}}, \bibinfo {author}
  {\bibfnamefont {L.~N.}\ \bibnamefont {Pfeiffer}}, \bibinfo {author}
  {\bibfnamefont {K.~W.}\ \bibnamefont {Baldwin}}, \ and\ \bibinfo {author}
  {\bibfnamefont {K.~W.}\ \bibnamefont {West}},\ }\href {\doibase
  10.1103/PhysRevLett.93.176809} {\bibfield  {journal} {\bibinfo  {journal}
  {Phys. Rev. Lett.}\ }\textbf {\bibinfo {volume} {93}},\ \bibinfo {pages}
  {176809} (\bibinfo {year} {2004})}\BibitemShut {NoStop}%
\bibitem [{\citenamefont {Pan}\ \emph {et~al.}(2008)\citenamefont {Pan},
  \citenamefont {Xia}, \citenamefont {Stormer}, \citenamefont {Tsui},
  \citenamefont {Vicente}, \citenamefont {Adams}, \citenamefont {Sullivan},
  \citenamefont {Pfeiffer}, \citenamefont {Baldwin},\ and\ \citenamefont
  {West}}]{pan2008}%
  \BibitemOpen
  \bibfield  {author} {\bibinfo {author} {\bibfnamefont {W.}~\bibnamefont
  {Pan}}, \bibinfo {author} {\bibfnamefont {J.~S.}\ \bibnamefont {Xia}},
  \bibinfo {author} {\bibfnamefont {H.~L.}\ \bibnamefont {Stormer}}, \bibinfo
  {author} {\bibfnamefont {D.~C.}\ \bibnamefont {Tsui}}, \bibinfo {author}
  {\bibfnamefont {C.}~\bibnamefont {Vicente}}, \bibinfo {author} {\bibfnamefont
  {E.~D.}\ \bibnamefont {Adams}}, \bibinfo {author} {\bibfnamefont {N.~S.}\
  \bibnamefont {Sullivan}}, \bibinfo {author} {\bibfnamefont {L.~N.}\
  \bibnamefont {Pfeiffer}}, \bibinfo {author} {\bibfnamefont {K.~W.}\
  \bibnamefont {Baldwin}}, \ and\ \bibinfo {author} {\bibfnamefont {K.~W.}\
  \bibnamefont {West}},\ }\href {\doibase 10.1103/PhysRevB.77.075307}
  {\bibfield  {journal} {\bibinfo  {journal} {Phys. Rev. B}\ }\textbf {\bibinfo
  {volume} {77}},\ \bibinfo {pages} {075307} (\bibinfo {year}
  {2008})}\BibitemShut {NoStop}%
\bibitem [{\citenamefont {Kumar}\ \emph {et~al.}(2010)\citenamefont {Kumar},
  \citenamefont {Cs\'athy}, \citenamefont {Manfra}, \citenamefont {Pfeiffer},\
  and\ \citenamefont {West}}]{kumar2010}%
  \BibitemOpen
  \bibfield  {author} {\bibinfo {author} {\bibfnamefont {A.}~\bibnamefont
  {Kumar}}, \bibinfo {author} {\bibfnamefont {G.~A.}\ \bibnamefont {Cs\'athy}},
  \bibinfo {author} {\bibfnamefont {M.~J.}\ \bibnamefont {Manfra}}, \bibinfo
  {author} {\bibfnamefont {L.~N.}\ \bibnamefont {Pfeiffer}}, \ and\ \bibinfo
  {author} {\bibfnamefont {K.~W.}\ \bibnamefont {West}},\ }\href {\doibase
  10.1103/PhysRevLett.105.246808} {\bibfield  {journal} {\bibinfo  {journal}
  {Phys. Rev. Lett.}\ }\textbf {\bibinfo {volume} {105}},\ \bibinfo {pages}
  {246808} (\bibinfo {year} {2010})}\BibitemShut {NoStop}%
\bibitem [{\citenamefont {Nadj-Perge}\ \emph {et~al.}(2014)\citenamefont
  {Nadj-Perge}, \citenamefont {Drozdov}, \citenamefont {Li}, \citenamefont
  {Chen}, \citenamefont {Jeon}, \citenamefont {Seo}, \citenamefont {MacDonald},
  \citenamefont {Bernevig},\ and\ \citenamefont {Yazdani}}]{nadj-perge2014}%
  \BibitemOpen
  \bibfield  {author} {\bibinfo {author} {\bibfnamefont {S.}~\bibnamefont
  {Nadj-Perge}}, \bibinfo {author} {\bibfnamefont {I.~K.}\ \bibnamefont
  {Drozdov}}, \bibinfo {author} {\bibfnamefont {J.}~\bibnamefont {Li}},
  \bibinfo {author} {\bibfnamefont {H.}~\bibnamefont {Chen}}, \bibinfo {author}
  {\bibfnamefont {S.}~\bibnamefont {Jeon}}, \bibinfo {author} {\bibfnamefont
  {J.}~\bibnamefont {Seo}}, \bibinfo {author} {\bibfnamefont {A.~H.}\
  \bibnamefont {MacDonald}}, \bibinfo {author} {\bibfnamefont {B.~A.}\
  \bibnamefont {Bernevig}}, \ and\ \bibinfo {author} {\bibfnamefont
  {A.}~\bibnamefont {Yazdani}},\ }\href {\doibase 10.1126/science.1259327}
  {\bibfield  {journal} {\bibinfo  {journal} {Science}\ }\textbf {\bibinfo
  {volume} {346}},\ \bibinfo {pages} {602} (\bibinfo {year}
  {2014})}\BibitemShut {NoStop}%
\bibitem [{\citenamefont {Pfeifer}\ and\ \citenamefont
  {Singh}()}]{pfeifer2015}%
  \BibitemOpen
  \bibfield  {author} {\bibinfo {author} {\bibfnamefont {R.~N.~C.}\
  \bibnamefont {Pfeifer}}\ and\ \bibinfo {author} {\bibfnamefont
  {S.}~\bibnamefont {Singh}},\ }\href {http://arxiv.org/abs/1505.00100v1}
  {}\Eprint {http://arxiv.org/abs/1505.00100v1}
  {arXiv:1505.00100v1 [cond-mat.str-el] (2015)} \BibitemShut {NoStop}%
\end{thebibliography}
\end{document}